\documentclass{aa}  

\usepackage[normalem]{ulem}
\usepackage{graphicx}
\usepackage{color}
\usepackage{url}
\usepackage[breaklinks,colorlinks,urlcolor=blue,citecolor=blue,linkcolor=blue]{hyperref}
\usepackage{lipsum}
\usepackage{multirow}

\usepackage{txfonts}
\usepackage{natbib}
\usepackage{amsmath}    % Advanced maths commands
\usepackage{amssymb} 
\usepackage{bm}
\usepackage{mathtools}
\usepackage{xcolor}
\newcommand{\SO}{MAXI~J1820$+$070}
\newcommand\hxmt{{\it Insight}-HXMT}
\newcommand\swift{{\it Swift}}
\definecolor{dodgerblue}{RGB}{30, 144, 255}

\begin{document} 

%%%%%%%%
\title{Evidence for enhanced mass transfer in the disc preceding the transition to the soft state in \SO}

\titlerunning{Enhanced mass transfer in the disc before the transition to the soft state}

\author{Pengcheng Yang\inst{1}\thanks{E-mail: pyang@astro.rug.nl}
\and
Guobao Zhang\inst{2,3}\thanks{E-mail: zhangguobao@ynao.ac.cn}
\and
David M. Russell\inst{4}
\and
Mariano M\'endez\inst{1}\thanks{E-mail: mariano@astro.rug.nl}
\and
M. Cristina Baglio\inst{5}
\and
Diego Altamirano\inst{6}
\and
Yijung Yang\inst{7,8}
\and
Payaswini Saikia \inst{4}
\and 
Kevin Alabarta \inst{4}}

\institute{Kapteyn Astronomical Institute, University of Groningen, PO Box 800, NL-9700 AV Groningen, The Netherlands
\and 
Yunnan Observatories, Chinese Academy of Sciences, Kunming 650216, People's Republic of China
\and 
Key Laboratory for the Structure and Evolution of Celestial Objects, Chinese Academy of Sciences, Kunming 650216, People’s Republic of China
\and
Center for Astrophysics and Space Science (CASS), New York University Abu Dhabi, PO Box 129188, Abu Dhabi, UAE
\and
INAF, Osservatorio Astronomico di Brera, Via E. Bianchi 46, I-23807 Merate (LC), Italy
\and
School of Physics and Astronomy, University of Southampton, Southampton, Hampshire SO17 1BJ, UK
\and
Graduate Institute of Astronomy, National Central University, 300 Zhongda Road, Zhongli, Taoyuan 32001, Taiwan
\and
Laboratory for Space Research, The University of Hong Kong, Cyberport 4, Hong Kong
}
 
\abstract
{We investigate the 2018--2019 main outburst and the subsequent mini-outbursts of the black hole low-mass X-ray binary \SO\ using optical/ultraviolet data from the Las Cumbres Observatory (LCO), the American Association of Variable Star Observers (AAVSO), and \swift/UVOT, as well as X-ray data from \hxmt\ and \swift/XRT. Given the high-cadence observations, we identify a broad dip-like feature in both the optical and X-ray light curves preceding the transition to the soft state, with the X-ray dip lagging the optical dip by approximately 10 days.
We propose that the dip is caused by a brief decrease followed by an increase in the mass accretion rate as it propagates through the disc, ultimately triggering the transition to the soft state. This might be a potential tool to predict impending hard-to-soft state transitions, although such a dip has not yet been observed in many sources.
Additionally, we find that optical colour ($g^{\prime}-i^{\prime}$) becomes bluer and less variable before the transition to the intermediate state, preceding a dramatic change in the hardness ratio. This appears to be an unusual case, differing from the typical scenario where the optical colour changes usually along with the transition to the soft state.
Finally, we explore the implications of the complex evolution of optical/X-ray correlation during both main outbursts and mini-outbursts. In particular, we find a loop-like evolutionary track before the transition to the soft state, which is linked to the optical/X-ray dips in the light curves.}

\maketitle

%-------------------------------------------------------------------

\section{Introduction}
\label{sec: introduction}
Black hole low-mass X-ray binaries (BH-LMXB) are systems in which a stellar-mass black hole (BH) accretes material from a non-degenerate companion star, typically with a mass of $\le$ 1 $M_\odot$, via an accretion disc surrounding the BH. Most BH-LMXBs are transient sources, exhibiting bright outbursts that last weeks to months, interspersed with quiescent periods that can persist for years \citep[e.g.,][]{1997ApJ...491..312C, 2016ApJS..222...15T, 2023hxga.book..120B}. These outbursts are believed to be driven by thermal-viscous instability in the accretion disc \citep[e.g.,][]{2001NewAR..45..449L}. During an outburst, the brightness of these transient systems scales with the mass accretion rate, making it several orders of magnitude brighter than in quiescence \citep[e.g.][]{1997MNRAS.292..679L, 2023hxga.book..120B}. 

BH-LMXBs emit radiation across the broadband electromagnetic spectrum during outbursts. The radio emission is believed to originate from synchrotron radiation of electrons in a relativistic, collimated outflow, or jet. The spectra of these systems from the ultraviolet (UV) to the near-infrared bands are more complex, as these wavelengths lie at the intersection of multiple competing emission mechanisms. For instance, the X-ray reprocessing either by the companion star surface or by the outer accretion disc, viscous heating from the outer accretion disc, synchrotron emission from the jet, or synchrotron radiation in a hot flow \citep{1973A&A....24..337S, 1976ApJ...208..534C, 1994A&A...290..133V, 2006MNRAS.371.1334R, 2013MNRAS.430.3196V}. The X-ray spectrum is typically attributed to two main components: a thermal component and a non-thermal component. The thermal component is thought to originate from a geometrically thin, optically thick accretion disc \citep{1973A&A....24..337S}, which can be described by a multi-colour disc blackbody \citep{1984PASJ...36..741M}, with a peak blackbody temperature varying between 0.1 keV and 2.5 keV \citep{1997AIPC..410..141Z}. The non-thermal component is generally thought to arise from a hot electron plasma, often referred to as the corona, through inverse Compton scattering of the soft photons from the accretion disc or synchrotron emission of the relativistic electrons \citep[e.g.][]{1979ApJ...229..318G, 2018A&A...614A..79P}. This component can be characterised by a power-law with an exponential cut-off at tens to hundreds of keV \citep[e.g.][]{1979Natur.279..506S, 1980A&A....86..121S}. Despite decades of numerous studies, there is still no consensus on the detailed geometry of the corona.

Most BH-LMXBs exhibit almost consistent behaviour during outbursts, tracing an anti-clockwise ``q'' shape in the hardness-intensity diagram \citep[e.g.,][]{2001ApJS..132..377H, 2006MNRAS.367.1113B}. Based on distinct X-ray spectral and variability properties, two commonly used state classification schemes are proposed to describe the outburst evolution: one by \citet{2005Ap&SS.300..107H} and another by \citet{2006ARA&A..44...49R}. The former emphasises on the transitions between states. \citep[see also][for review]{2016ASSL..440...61B}, while the latter put more emphasis on
the ``stable'' states. In this work, I will adopt definitions from both schemes. At the beginning of the outburst, the source is in the hard state (HS), where the X-ray spectrum is dominated by Comptonised power-law emission from the corona. During this phase, a compact jet can be usually detected.
As the luminosity increases, the X-ray spectrum softens as the hard corona flux decreases and the soft disc flux increases. When the thermal and non-thermal spectral components contribute comparably, the source enters the intermediate state (IMS), at the end of which one or more relativistic and discrete jets are launched \citep[e.g.][]{2004MNRAS.355.1105F}. The IMS can be further subdivided into a hard intermediate state (HIMS) and a soft intermediate state (SIMS). The evolution continues until the thermal disc dominates the X-ray spectrum, coinciding with the jet quenching and, in some sources, the emergence of the X-ray wind. At this stage, the system is said to be in the soft state (SS). Eventually, as the outburst declines, the source goes back to the HS as it returns to quiescence.

The X-ray variability features that indicate different states can be described and quantified by the properties of the power density spectrum (PDS) \citep[see][for review]{2010LNP...794...53B}. Here, we briefly summarise the types of quasi-periodic oscillations (QPOs) and the  root-mean-square (rms) amplitude of the PDS in different states. There is a strong rapid X-ray variability with typical rms values of approximately 30\% in the HS. During the HIMS, the rms decreases to around 10--20\%. Type-C low-frequency QPOs usually emerge at the end of HS and during HIMS. At the onset of SIMS, Type-B QPOs replace the Type-C QPOs \citep[e.g.][]{2023MNRAS.525..854M}. The variability in this state can be as low as a few percent. In the SS, the rapid X-ray variability is much weaker, with rms values of $\lesssim$ 1\%. 
 
Most BH-LMXB's outbursts undergo transitions between these states, known as full outbursts, while some outbursts remain confined to the HS or reach only up to the HIMS \citep[e.g][]{2016ApJS..222...15T, 2021MNRAS.507.5507A}, called ``failed-transition'' (FT) outbursts. The precise physical mechanism triggering the state transition remains an open question. The state transition process typically occurs rapidly, leaving limited observational constraints. However, multi-wavelength observations have presented several clues that may aid in predicting state transitions. 
\citet[e.g.][]{2004ApJ...603..231K} found that during soft-to-hard state transitions, changes in X-ray timing properties (rms amplitude) are more pronounced than changes in the spectral properties, making them a more effective indicator of state transitions---although they lag behind spectral evolution.
\citet{2015ApJ...808..122F,2021MNRAS.507.5507A, 2021MNRAS.502..521D} observed a potential difference between full and FT outbursts in the radio/X-ray correlation. 
\citet{2021MNRAS.507.5507A} explored the observational difference between full and FT outbursts of GX 339--4 in the optical/infrared (IR) bands, finding that GX 339--4 appears brighter at optical/IR wavelengths before the onset of a FT outburst than before the onset of a full outburst. \citet{2023ApJ...958..153L} reported that in some BH-LMXBs, the X-ray variability, quantified by the power spectral hue \citep[][]{2015MNRAS.448.3348H, 2015MNRAS.448.3339H}, behaves in a systematic way, evolving a few weeks ahead of the hard-to-soft state transition quantified by the spectral hardness. It is also known that higher-energy emission from the compact jet, which in some sources radiates at IR or higher frequencies, quenches at the start of the state transition \citep[][]{2005ApJ...624..295H, 2007MNRAS.379.1401R, 2009MNRAS.400..123C, 2018ApJ...867..114B}, although there are instances of the appearance of a compact jet in the soft-intermediate state \citep{2020MNRAS.495..182R}. Additionally, \citet{2012MNRAS.427L..11Y} found a UV flux decrease preceding the HS-to-HIMS transition, which could signify the jet quenching at higher frequencies before the lower frequencies, throughout the state transition.

\SO\ is an X-ray binary initially discovered as an optical transient (ASASSN-18ey) by the All-Sky Automated Survey for SuperNovae (ASAS-SN) on UT 2018 March 06.58 (MJD 58184)\footnote{\url{http://www.astronomy.ohio-state.edu/asassn/transients.html}}
\citep[][]{2018ApJ...867L...9T}. Several days later, on 2018 March 11 (MJD 58188), Monitor of All-sky X-ray Image (MAXI) detected a bright X-ray transient (\SO) at the same location, later identified as \SO\ \citep[][]{2018ATel11399....1K}. Follow-up multiwavelength observations confirmed that \SO\ is a black hoe X-ray binary \citep[e.g.][]{2018ATel11418....1B, 2018ATel11420....1B, 2018ATel11423....1U}. \SO\ is considered a high-inclination system, with an inclination of $i$ = $63 \pm 3^{\circ}$ \citep{2020MNRAS.493L..81A}. The mass of BH has been estimated to be $M_{\rm BH}=8.48^{+0.79}_{-0.72} M_{\odot}$ \citep{2020ApJ...893L..37T}. Based on radio parallax measurements, \citet{2020MNRAS.493L..81A} derive a distance of 2.96 $\pm$ 0.33 kpc to the source. Optical spectroscopic observations determined an orbital period of 16.4518 $\pm$ 0.0002 h \citep{2019ApJ...882L..21T}.
The main outburst in 2018 remains bright in multiwavelength for $\sim$ 200 days, undergoing full state transitions between the HS and SS \citep[e.g.][]{2019ApJ...874..183S}, before decaying close to the quiescence \citep[][]{2019ATel12534....1R}. Subsequently, the source experienced three evident rebrightenings in multiwavelength without state transition: the first \citep[][]{2019ATel12596....1B, 2019ATel12573....1B, 2019ATel12567....1U, 2019ATel12577....1W} and second \citep[][]{2019ATel13014....1H, 2019ATel13025....1X} in 2019, and the third \citep[][]{2020ATel13502....1A, 2020ATel13530....1S} in 2020. After several years of low-level activities, the latest reports indicate that the source continued to fade to lower level in optical, UV and X-ray band, and finally reached quiescence in optical band on 2023 August 5 (MJD 60161) \citep[][]{2023ATel16200....1H, 2023ATel16192....1B}.

We present a detailed multi-wavelength observations of the main outburst and following three mini-outbursts of \SO. The data and reduction methods used in this study are described in Sec. \ref{sec: Observations and Data Reduction}. The results, including the multi-wavelength light curves, X-ray hardness ratio, optical/UV colour and optical/X-ray flux correlation, are presented in Sec. \ref{sec: Results}
In Sec. \ref{sec: Discussion}, we discuss our findings, including the interpretation of the optical/UV colours during both the main outburst and mini-outbursts, the propagating optical/X-ray dips occurring before the hard-to-soft state transition, and the complex optical/X-ray correlations.

\begin{figure*}
\centering
    \mbox{\includegraphics[width=2.0\columnwidth]{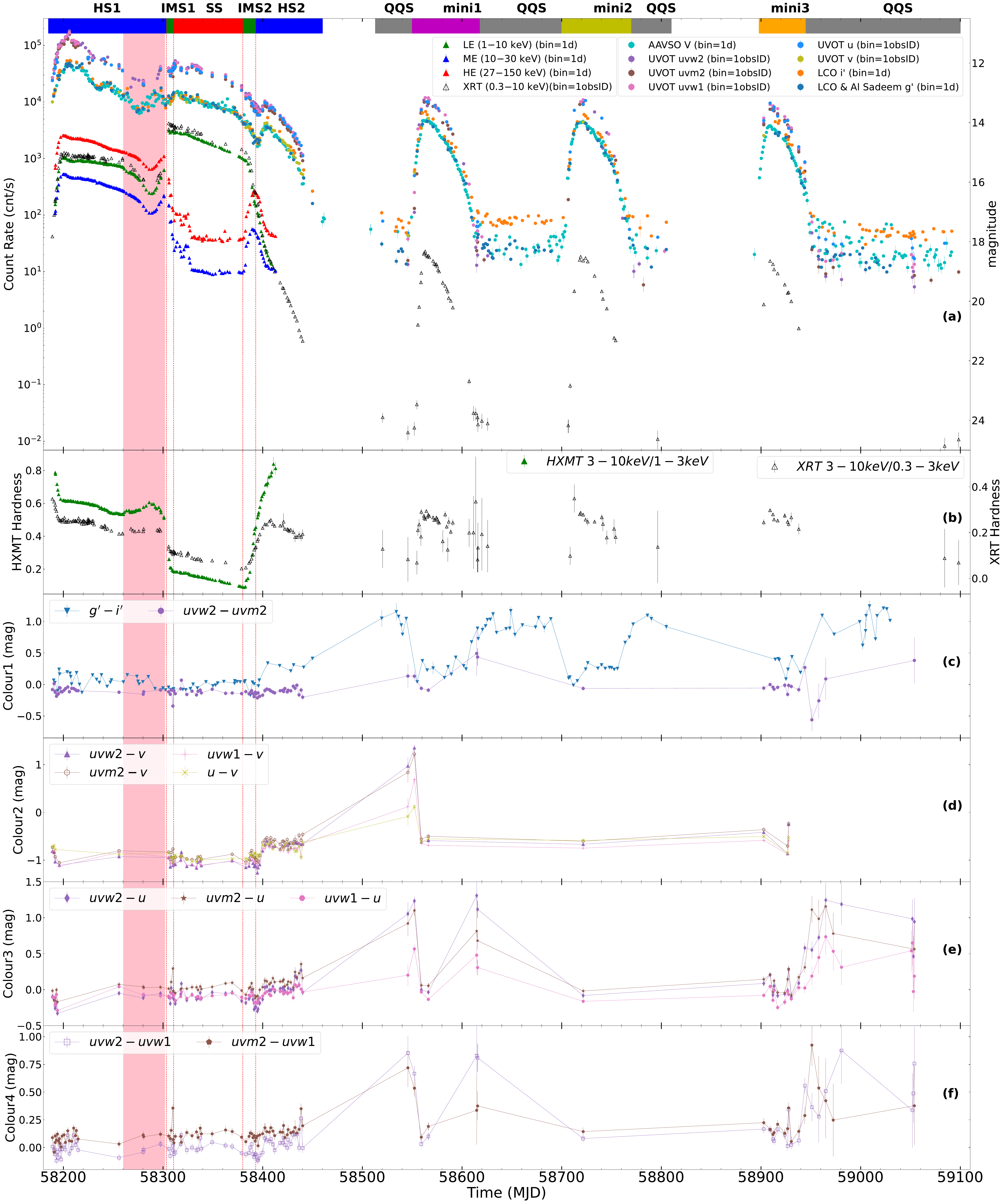}}
    \caption{Multi-wavelength light curves, X-ray hardness and optical/UV colours based on \hxmt, \swift\ and LCO, AAVSO observations of \SO. (a): Optical, UV and X-ray light curves. (b): Hardness ratios from \hxmt/LE and \swift/XRT. (c)--(f): Optical/UV colours defined by the magnitude difference between the optical or UV high-frequency and low-frequency bands. \textbf{Note:} (1) The different colours of bars at the top of panel (a) are used to differentiate the states of the main outburst, quasi-quiescent states and three mini-outbursts. ``mini1'', ``mini2'' and ``mini3'' represent the three mini-outbursts, respectively. The other acronyms are explained in the text. The QQS epochs are defined as periods when the $g^{\prime}$-band magnitude is fainter than 17.9. (2) The pink shaded region denotes the period of the optical and X-ray dip events. (3) The interval between the first two dashed red vertical lines at MJD $\sim$ 58303.5 and 58310.7 denotes the period of the hard-to-soft state transition. (4) The last two dashed red vertical lines at MJD $\sim$ 58380.0 and 58393.0 indicate the period of the soft-to-hard state transition.}
    \label{fig:time_evol_1}
\end{figure*}

\section{Observations and Data Reduction}
\label{sec: Observations and Data Reduction}

\subsection{\hxmt\ \& \swift /XRT: X-ray data}
The long-term and high-cadence observations of \SO\ by \hxmt\ spanned from 14 March 2018 (MJD 58191) to 21 October 2018 (MJD 58412), covering Exposure IDs P0114661001*--P0114661154*. \hxmt\ carries three instruments: the High Energy X-ray telescope (HE, 20--250 keV, \citealt{2020SCPMA..6349503L}), the Medium Energy X-ray telescope (ME, 5--30 keV, \citealt{2020SCPMA..6349504C}), and the Low Energy X-ray telescope (LE, 1--15 keV, \citealt{2020SCPMA..6349505C}). We extract light curves and spectra from these instruments using the \hxmt\ Data Analysis Software (HXMTDAS) v2.06\footnote{\url{http://hxmten.ihep.ac.cn/software.jhtml}}. In this study, we extract light curves in the 1--10 keV, 10--30 keV and 27--150 keV energy bands for the LE, ME and HE instruments, respectively, and rebin the light curves to 1-day bin size. We also calculate hardness ratios between the 1--3 keV and the 3--10 keV energy bands. 

Additionally, we collect archival X-ray data from the X-Ray Telescope (XRT) \citep{2005SSRv..120..165B} onboard \swift, including ObsIDs 00010627001 -- 00010627226, 00010754001 -- 00010754006, 0010774001 -- 00010774004, 00088779001, 00813771000, 00814259000 and 00815603000 (MJD 58189 -- 59098). Most of the XRT data we used are in the windowed-timing mode unless the observations are only in the photon-counting mode. Each observation has an average exposure time of a few kiloseconds. We use the \swift/XRT online product builder\footnote{\url{http://www.swift.ac.uk/user_objects/}} \citep{2009MNRAS.397.1177E} to extract a spectrum and the average photon count rate for each observation, obtaining the X-ray light curve in the 0.3--10 keV and the hardness between the 0.3--3 keV and 3--10 keV energy bands. 

To estimate the 2–10 keV X-ray fluxes used in the analysis of the optical/X-ray flux correlation, and focusing mainly on the emission from the continuum components, we adopt simplified spectral models to fit energy spectra of \hxmt\  and \swift/XRT. Our spectral fittings are conducted with \texttt{XSPEC} v12.15.0 \citep{1996ASPC..101...17A}. Using the  \texttt{grppha} tool, we group the \hxmt\ spectra with a minimum of 30 photons per bin and \swift/XRT Photon Counting (PC) or Windowed Timing (WT) mode spectra with a minimum of 20 photons per bin. Most of the \swift/XRT spectra we used are obtained in WT mode to avoid the pile-up effect present in PC mode. For the \hxmt\ data, we use the model \texttt{constant*TBabs*(diskbb+nthComp+gaussian)} to fit the spectra across the 2--150 keV energy range, using the LE (2--10 keV), ME (10--30 keV) and HE (28--150 keV) Telescopes. The \texttt{diskbb} component \citep{1984PASJ...36..741M} is used to characterise the multi-temperature blackbody component from the accretion disc, the \texttt{nthComp} component \citep{1996MNRAS.283..193Z, 1999MNRAS.309..561Z} to fit the Comptonisated spectrum due to inverse Compton scattering of thermal photons in the corona, and the \texttt{gaussian} component to account for the broadened iron emission line around 6.4 keV. The component \texttt{constant} represents the cross normalisation between the three instruments; we fix it to 1 for LE and let it free for ME and HE. We use the \texttt{Tbabs} component to account for interstellar absorption, with abundance tables from \citet{2000ApJ...542..914W} and cross section tables from \citet{1996ApJ...465..487V}, with the column density fixed at $N_{H}=0.15 \times 10^{22}$ cm$^{-2}$ \citep{2018ATel11423....1U}. A 1\% systematic error is added to the model. For the \swift/XRT\ data, we adopt \texttt{Tbabs*(diskbb+nthcomp)} to fit the spectra across the 0.3–10 keV energy range. We use the component \texttt{cflux} to calculate the 2--10 keV unabsorbed X-ray fluxes from the fitted \hxmt\ and \swift/XRT\ spectra, respectively, with $1\sigma$ errors. We note that, for some low-flux XRT observations where reliable spectral fitting is not possible, we estimate the upper limits of the X-ray fluxes by modelling a power-law spectrum ($\Gamma$ = 1.2) using the \texttt{pimms}\footnote{\url{https://heasarc.gsfc.nasa.gov/cgi-bin/Tools/w3pimms/w3pimms.pl}} tool.

\subsection{LCO, Al Sadeem and AAVSO: optical data}
\SO\ was monitored during its 2018 outburst and subsequent mini-outbursts at optical wavelengths by the Las Cumbres Observatory (LCO) telescopes, as part of an ongoing monitoring campaign of $\sim 50$ LMXBs coordinated by the Faulkes Telescope Project \citep[][]{2008arXiv0811.2336L}. Observations were carried out at {the} 1-m and 2-m (Faulkes) nodes of LCO. The X-ray Binary New Early Warning System (XB-NEWS) data analysis pipeline \citep[][Alabarta et al. in prep]{2019AN....340..278R, 2020MNRAS.498.3429G} was used for computing an astrometric solution for each image using Gaia DR2 positions, performing aperture photometry of all the stars in the image, solving for zero-point calibrations between epochs \citep[][]{2012MNRAS.424.1584B}, and flux calibrating the photometry using the ATLAS All-Sky Stellar Reference Catalog \citep{2018ApJ...867..105T}. Part of the LCO data set on specific dates was included in multi-wavelength spectral analyses in \citet{2024ApJ...962..116E} and \citet{2024ApJ...964..189B}, and will be included in Baglio et al. (in preparation). See \citet{2024ApJ...962..116E} for more details of the LCO analysis procedures. Here, we use the long-term LCO monitoring data in the $SDSS$ $g^{\prime}$ and $i^{\prime}$ bands (MJD 58190--58091). We discard the optical observations with magnitude errors > 0.25 mag, as these are non-detections or low-significance detections. We set the bin size of the light curve data to 1 day. 

The source was also monitored with the Al Sadeem Observatory\footnote{\url{http://alsadeemastronomy.ae/}}. The observations were conducted using a Meade LX850 16-inch (41-cm) telescope equipped with an SBIG STT-8300 camera, along with Baader LRGB CCD filters (blue, green and red filters with similar central wavelengths to $g^{\prime}$, $V$ and $R$-bands). For the details of the data analysis procedures, see \citet{2024ApJ...962..116E}. In this paper, the $g^{\prime}$-band light curve also includes a few monitoring data from Al Sadeem. The full light curves of the outburst of \SO\ from Al Sadeem will be published in Baglio et al. in preparation.

We also include high-cadence V-band observations (MJD 58193--59091) from the American Association of
Variable Star Observers (AAVSO)\footnote{\url{https://aavso.org/aavso-international-database-aid}}. We selected only observations with an error < 0.25 mag, resulting in a total of 184,577 data points, and averaged the light curve to a 1-day bin size.

To facilitate correlation analysis, we convert $g^{\prime}$ and V--band magnitude to flux densities, applying extinction corrections using the \citet{1999PASP..111...63F} extinction curve. We use the colour excess $E(B-V)=0.16$ \citep{2022MNRAS.514.3894O}, and adopt the Galactic extinction law with $R_V=A_V/E(B-V)=3.1$.
Using these parameters, we calculate the dereddened flux densities via the tool \texttt{unred}\footnote{\url{https://pyastronomy.readthedocs.io/en/latest/pyaslDoc/aslDoc/unredDoc.html}}  in PyAstronomy, which implements the \citet{1999PASP..111...63F} parametrization.

\subsection{\swift /UVOT: Optical \& Ultraviolet data}
\SO\ was simultaneously observed by the Ultraviolet/Optical Telescope (UVOT) onboard \swift\ in three UV ({\it uvw2}, {\it uvm2} and {\it uvw1}) and three optical filters ({\it u}, {\it b} and {\it v}). 
The {\it b}-band observations are only available for a few ObsIDs (00813771000, 00814259000, 00815603000).
As a result, in this study, we use all archival \swift/UVOT monitoring data except for the {\it b}-band data.
We use the HEAsoft task \texttt{uvotimsum} to sum the sky images for those observations with more than one image extension. To determine the flux density and magnitude of each observation using the \texttt{uvotsource}, we select a circular region with a radius of 5 arcsec centred on the source, and a circular source-free region with a radius of 20 arcsec for the background subtraction. The UV/optical photometry errors are determined at a $1\sigma$ confidence level. We obtain the UVOT light curve, in which each data point represents an observation ID.

\section{Results}
\label{sec: Results}

\subsection{Overall evolution of the light curves}
\label{subsec: overall_lc}
Fig.~\ref{fig:time_evol_1} (a) displays the \hxmt\ light curves of \SO\ in the X-ray band (\hxmt\ and XRT), optical band (AAVSO/V, LCO/$g^{\prime}$, LCO/$i^{\prime}$, Al Sadeem/$g^{\prime}$, {\it u} and {\it v}), and UV band ({\it uvw2}, {\it uvm2} and {\it uvw1}).
It shows that \SO\ underwent a main outburst followed by three mini-outbursts in the multi-wavelength light curves. 
We refer to these observations before MJD$\sim$58460 as the main outburst. At the beginning of the main outburst phase, \SO\  exhibits a rapid rise to its first peak across the optical, UV, and X-ray bands, with the optical and UV peaks coming approximately 5–6 days later than the X-ray peak. After the first peak, the source experiences a gradual decay without reaching quiescence, followed by two rebrightenings in the optical, UV, and X-ray bands, occurring around MJD 58290 and 58390. The first one appears just before the transition to the IMS. The second one, in the optical/UV bands, takes place shortly after the transition from soft to hard state, and it also occurs during the soft-to-hard transition in the hard X-ray band. The source stays in the soft state between these two rebrightenings. After the main outburst, three additional rebrightening events are observed in the optical, UV and X-ray bands. 
Between the main outburst and the last three rebrightenings, faint observations are seen, with the average optical magnitude ($g^{\prime}$=18.43) still brighter than the real quiescence ($g^{\prime}$=19.64, \cite{2023ATel16192....1B}). We define these faint observations, where the $g^{\prime}$-band magnitude is fainter than 17.9, as a quasi-quiescent state (QQS).
We approximately treat the quasi-quiescent state as the quiescent state. Based on the three criteria: (1) the source flux reaches quiescence before the rebrightening (2) the flux ratio between each rebrightening peak flux and the outburst peak flux is < 0.7; (3) the ratio of the quiescent period before rebrightening to the duration of the main outburst  $\leq$ 1 \citep[see][for more details of the classification]{2019ApJ...876....5Z}, we classify the last three rebrightenings as mini-outbursts, although here the start of the quasi-quiescence after the main outburst can't be determined well.
The light curves of the mini-outbursts show a rapid rise, followed by a slightly slower decay. All three mini-outbursts have similar peak brightness and duration.
In addition, the mini-outbursts' optical/UV peak brightness nearly reaches that of the second rebrightening of the main outburst.

\subsection{X-ray hardness ratio}
\label{subsec: oval evo hard}
Fig.~\ref{fig:time_evol_1} (b) shows the evolution of the hardness ratio as time. The changes in the hardness ratio clearly indicate the source state.
We apply the definition of the spectral states of \citet{2019ApJ...874..183S} to this study. Before MJD 58303.5 and during MJD 58393.0--58460.0, the source is in the HS (hereafter referred to as HS1 state and HS2, respectively). During MJD 58303.5--58310.7 and 58380.0--58393.0, the source is in the IMS (hereafter referred to as IMS1 state and IMS2, respectively). The source is in the SS between MJD 58310.7--58380.0.
For three mini-outbursts, compared to the hardness values during the main outburst, in all the mini-outbursts the source remains in the hard state.

\begin{figure}
\centering
    \mbox{\includegraphics[width=\columnwidth]{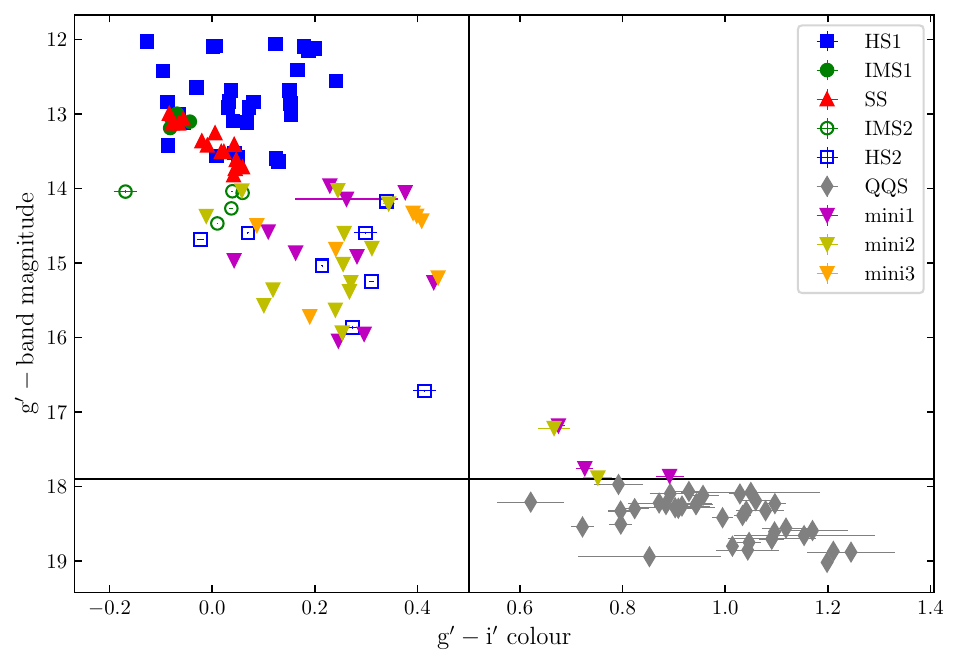}}
    \caption{Colour-magnitude diagram of \SO. The correlation between $g^{\prime}$-band magnitude and the $g^{\prime}-i^{\prime}$ colour. The horizontal line represents the threshold of quasi-quiescence (i.e. fainter than 17.9). The vertical line indicates the division of optical colour between the QQS period and the periods of the main outburst and three mini-outbursts. Data points in this figure are colour-coded according to the states defined in Fig. \ref{fig:time_evol_1}.}
    \label{fig:cmd}
\end{figure}

\begin{figure}
\begin{center}
	\includegraphics[width=1.0\columnwidth]{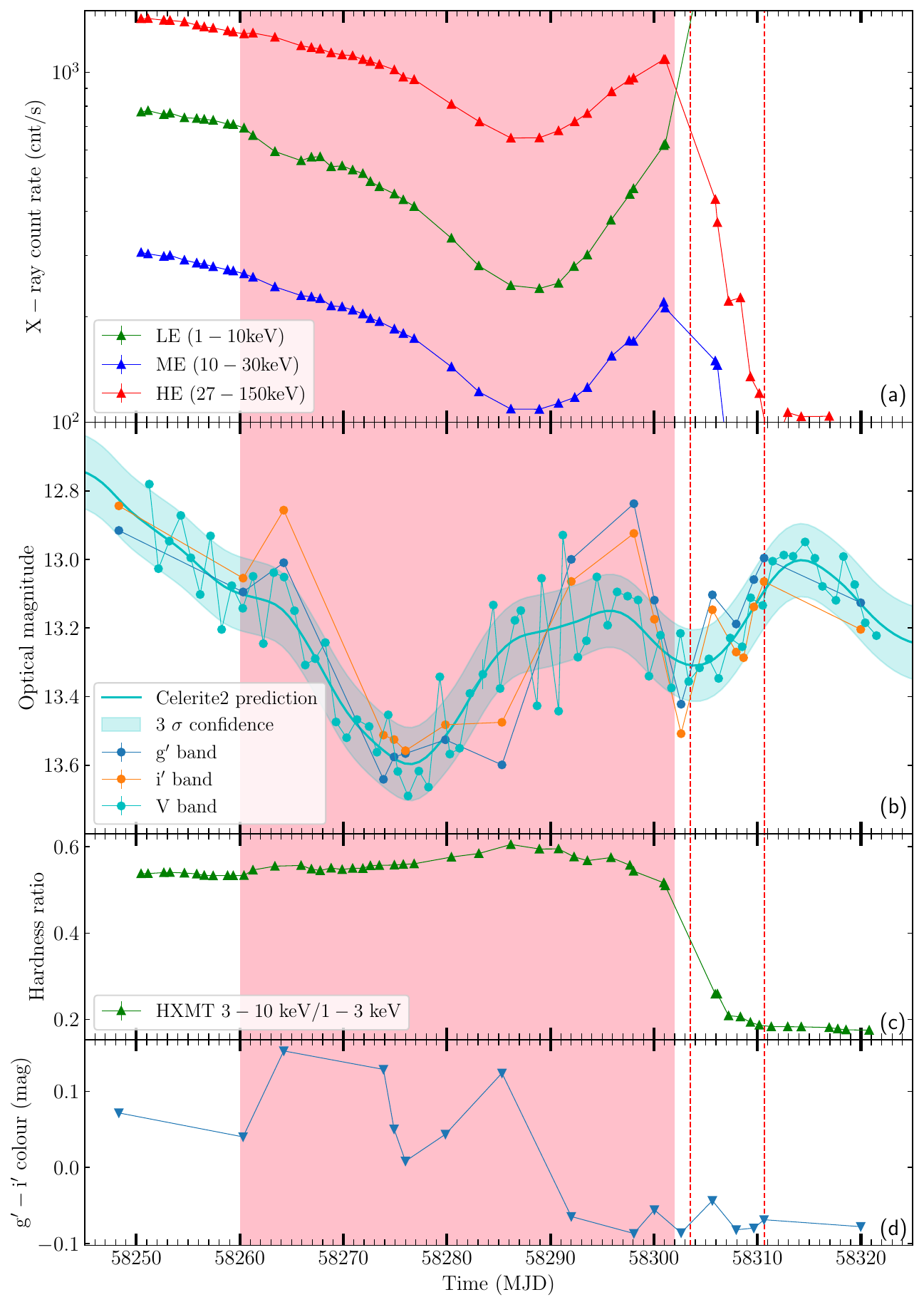}
    \caption{Optical/X-ray light curves, 
    hardness ratio and optical colour evolution during the dip events. (a)  X-ray light curves from \hxmt; (b) $g^{\prime}$, $i^{\prime}$, V-band light curves and 3$\sigma$ confidence curve of \texttt{Celerite2} for V-band light curve; (c) HXMT hardness; (d)  $g^{\prime}-i^{\prime}$ colour. The pink area represents the epoch of optical and X-ray dips. The interval between the two dashed red vertical lines marks the first intermediate-state period.}
    \label{fig:x_opt_lc_hardness_colour}
\end{center}
\end{figure}

\begin{figure*}
\begin{center}
	\includegraphics[width=2.0\columnwidth]{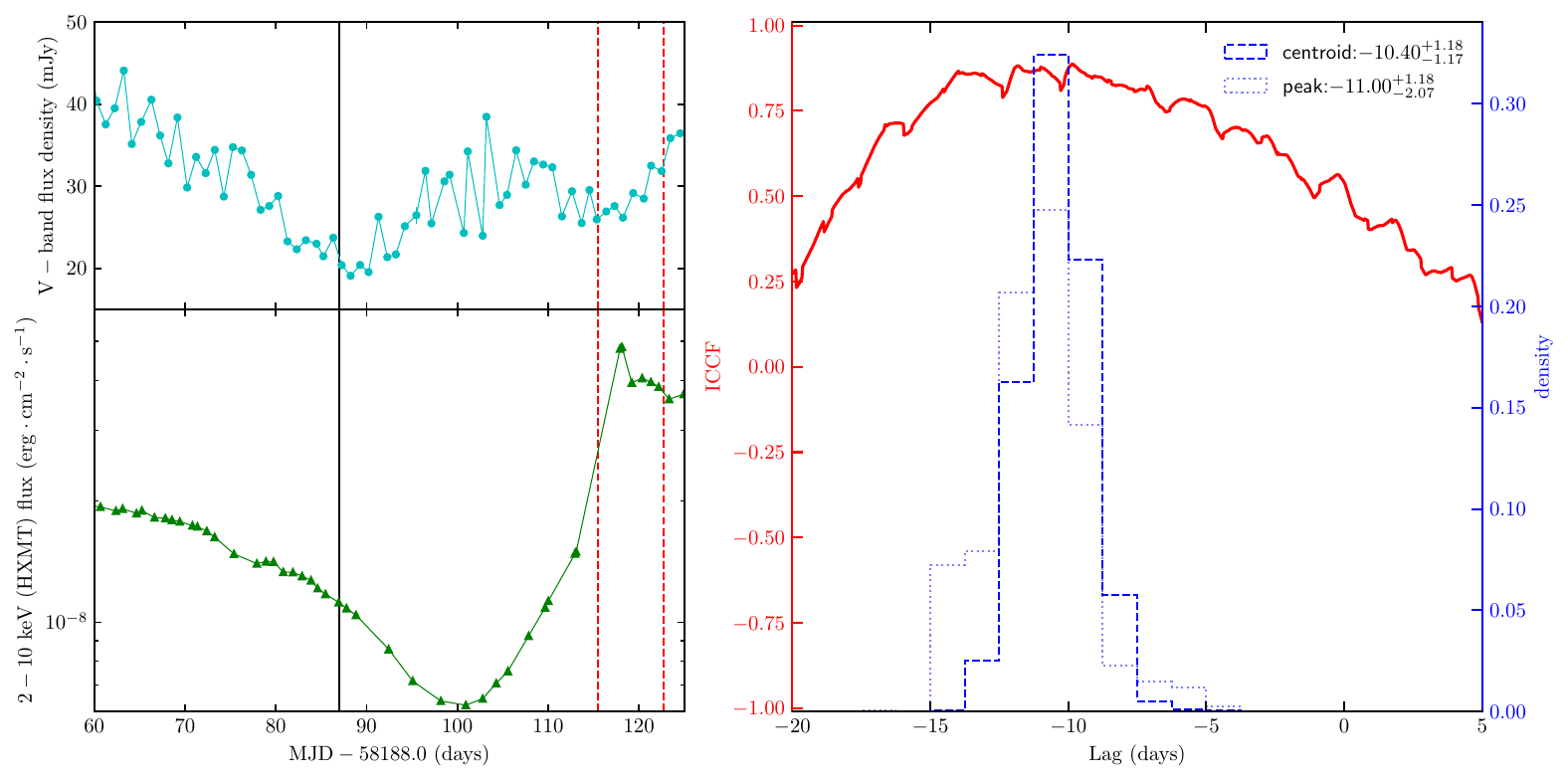}
    \caption{Cross correlation analysis between the optical (V band) and 2--10 keV X-ray (\hxmt) light curves of \SO. Left panel: Dip events in the X-ray and optical light curves. \textbf{Note}: (1) For convenient reference to the period of the occurrence of optical superhump discussed in \citet{2022MNRAS.509.1062T} and the X-ray/optical dips, we adopt the start date as used by them (i.e. day0 is MJD 58188.0). (2) The solid black vertical line (day 87) marks the onset of large-amplitude optical modulation at the superhump period. 
    The interval between the two dashed red vertical line is the intermediate state. Right: The red line shows the cross-correlation function (left axis). The blue dashed and dotted histograms represent the cross-correlation centroid lag distribution and the peak lag distribution, respectively (right axis). }
    \label{fig:HXMT-aavso}
\end{center}
\end{figure*}

\subsection{Overall evolution of the optical and UV colours}
\label{subsec: overall_colours}

Fig.~\ref{fig:time_evol_1} (c)--(f) show different colours variations derived from optical or UV data from LCO, Al Sadeem and UVOT. All colours are defined as the difference between the apparent magnitudes within 1 day, with the shorter wavelength minus the longer wavelength.
Among these, the $g^{\prime}-i^{\prime}$ colour in Fig.~\ref{fig:time_evol_1} (c) provides a better coverage than other colours for tracking the overall evolution. 

Colour-magnitude diagram (CMD) is a useful tool for inferring disc temperatures and disentangling jet and disc emission \citep[e.g.][]{2008ApJ...688..537M, 2011MNRAS.416.2311R}.
The disc is expected to appear bluer when brighter. If there is a jet present, that can make the data redder than the disc. 
Fig.~\ref{fig:cmd} shows the CMD of \SO\ throughout the main outburst, three mini-outbursts and QQS. The entire data set follows a general trend in which the optical colour becomes redder as it fades. It also reveals that the $g^{\prime}-i^{\prime}$ colour values during the mini-outburst are similar to the HS2 (between 0 and 0.4) of the main outburst, but it is fainter and slightly redder than in the HS1 and SS. 
In contrast, during the quasi-quiescence, the $g^{\prime}-i^{\prime}$ colour values are higher, ranging between $\sim$ 0.6 and 1.2. In the soft state, it shows a clear negative correlation between the optical colour and the optical magnitude (i.e. the optical colour becomes redder as the source gradually becomes fainter), which is expected for the emission from the accretion disc. During the hard state, the CMD displays a scattered distribution redder than the soft-state track.

Fig.~\ref{fig:time_evol_1} (d)--(f) show the other UV/optical colours consisting of \swift/UVOT data. The coverage of UVOT colour is primarily during the main outburst period, which will be mentioned in detail in Sec. \ref{subsec: main_outburst_colours}. We noticed that in the third mini-outburst, the UV colour becomes redder as it fade to the quasi-quiescence, which is similar to the behaviour in the optical ($g^{\prime}-i^{\prime}$) colour when entering QQS.

\subsection{Colour evolution during the main outburst}
\label{subsec: main_outburst_colours}

In the third panel in Fig.~\ref{fig:time_evol_2}, from the relatively good coverage of the optical colour ($g^{\prime}-i^{\prime}$) colour, 
we observe that the variability of the optical colour is stronger in the HS than the SS. This phenomenon is also reflected in the scattering during the hard state in the CMD (see Fig. \ref{fig:cmd}). From Fig. \ref{fig:x_opt_lc_hardness_colour} (c) and (d), we notice that the optical colour suddenly decline to reaching a lower value and being less variable before evolving into intermediate state, meanwhile, the X-ray hardness ratio shows only minor changes. In contrast, the prominent drop in the hardness ratio appears afterward, around the time of transition to soft state.

During the period of the soft-to-hard state transition in Fig.~\ref{fig:time_evol_2} (c)--(f), especially in last three groups of colours, the evolution of the colours consistently exhibits a pattern where a ``colour bump'' appears during the transition period (between the last two dashed red vertical lines in Fig.~\ref{fig:time_evol_2}). After entering the second hard state, a stronger ``colour bump'' appears.

\begin{figure}
\center
    \mbox{\includegraphics[width=1\columnwidth]{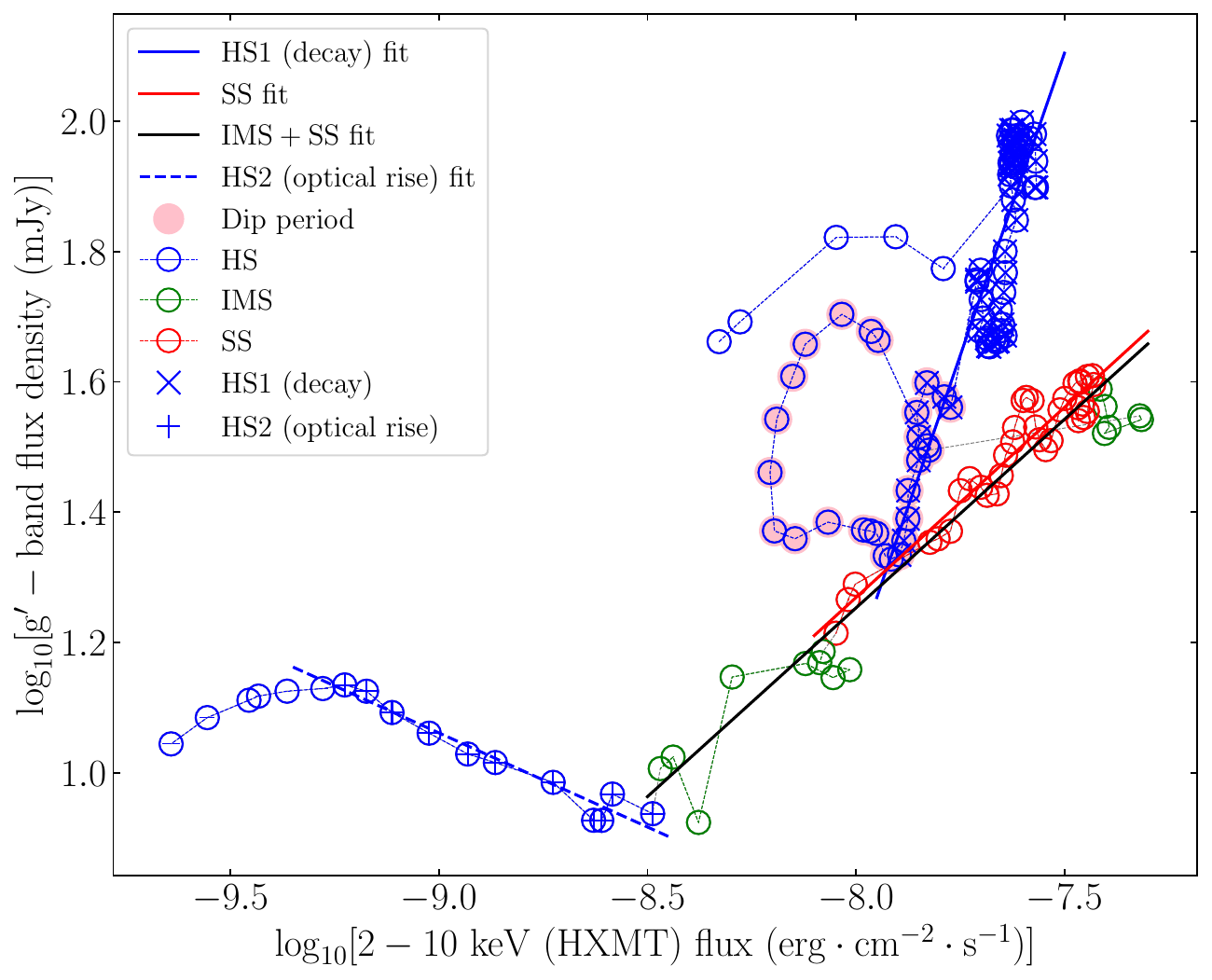}}
    \caption{Correlation between the 2-10 keV X-ray flux (\hxmt) and the interpolated optical ($g^{\prime}$ band) flux density during the main outburst in \SO.}
    \label{fig:g-hxmt_correlation_interpo}
\end{figure}

\begin{figure*}
\begin{center}
    \includegraphics[width=2\columnwidth]{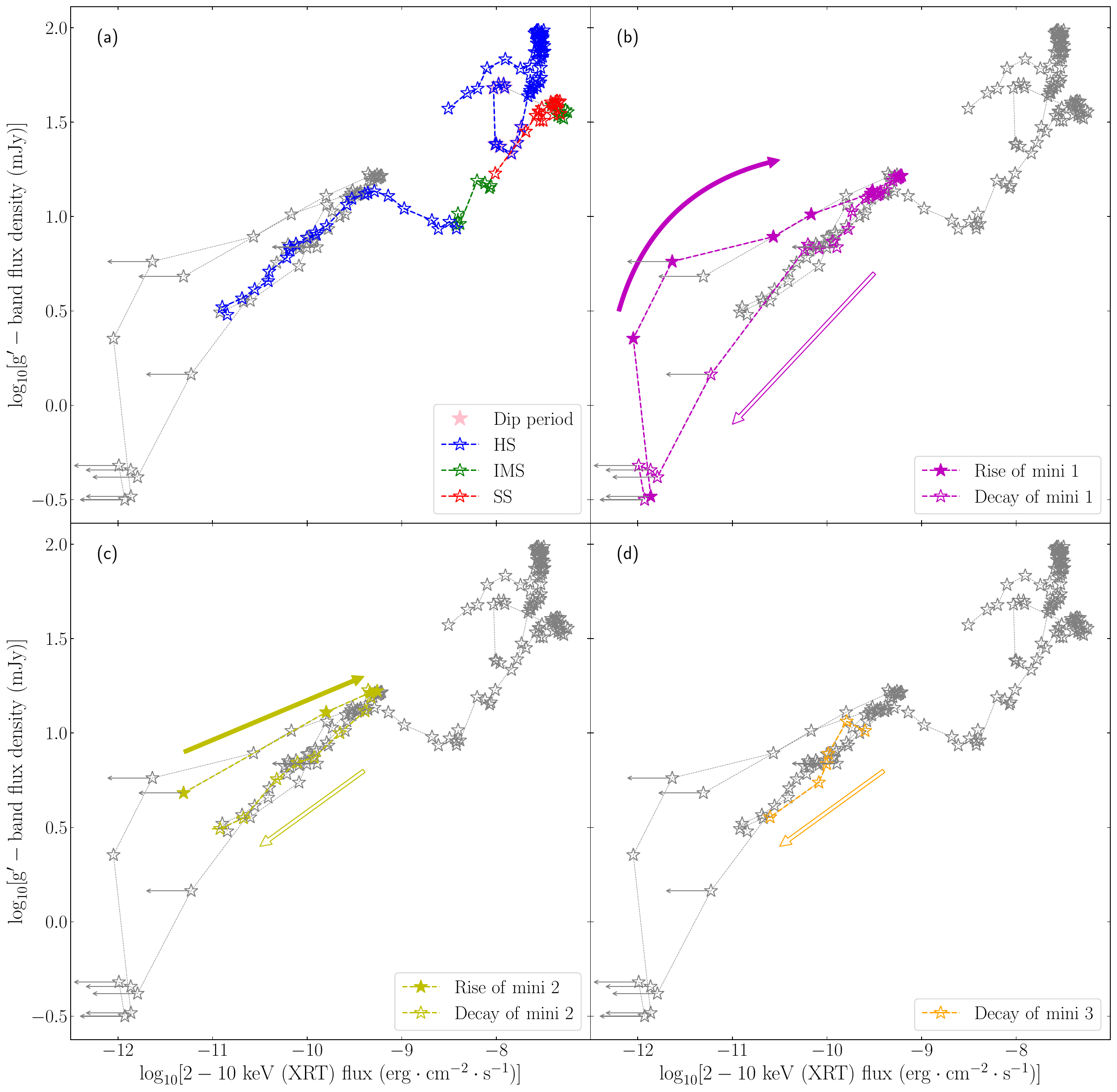}
    \caption{Correlation between the 2-10 keV X-ray flux (\swift/XRT) and the interpolated optical ($g^{\prime}$-band) flux density in \SO. The four panels show the main outburst and three mini-outbursts in the correlation diagram. Data points with gray arrows indicate upper limits.}
    \label{fig:g-xrt_correlation_interpo}
\end{center}
\end{figure*}

\subsection{Broad dip event before the hard-to-soft transition}
\label{subsec: broad_dip}

As shown in Fig. \ref{fig:time_evol_2} (a), we can see that the flux of LE, ME and HE decrease slowly until MJD$\sim$58260. Afterward, the flux start to decrease faster until MJD$\sim$58290 where the trend reverses, and the flux increases back to the level observed at MJD$\sim$58260, forming a dip in the X-ray light curves. The same behaviour is observed in the optical (the UV is less well constrained), especially in the AAVSO V-band monitoring. In V-band light curve, however, the dip appears to start (MJD$\sim$58250) earlier than in the X-ray band, reaching its minimum earlier (MJD$\sim$58275) before returning to the level observed at MJD$\sim$58250. From now on, we refer to this phenomenon as the dip observed in the X-ray and V-band light curves, as marked by the pink shaded area in Fig. \ref{fig:x_opt_lc_hardness_colour}. Fig. \ref{fig:x_opt_lc_hardness_colour} (a) and (b) provides a clearer view of the optical and X-ray dip events, focusing specifically on this phase. In both the X-ray and V-bands, the dip lasts for approximately 40 days. We note that the dip events occur prior to the transition from the hard to soft state.

A delay is observed, with the X-ray dip occurring later than the optical/UV dip. 
The $g^{\prime}$ and $i^{\prime}$-band light curves show a broad dip structure, but the fall and rise that we see with LCO may be misleading, as they could simply reflect different phases of the 0.5-mag amplitude modulation reported by \citet{2022MNRAS.509.1062T}. The monitoring in the UV band, on the other hand, lacks coverage during the dip events. The AAVSO, however, provides good sampling in the V band, making it suitable for analysing the delay between the X-ray dip and the V-band dips.
\citet{2022MNRAS.509.1062T} reported that the V-band light curve displays a large modulation, but the normalisation of the modulation changes around the epoch of the optical dip. We also find a clear dip structure (see the prediction curve in Fig. \ref{fig:HXMT-aavso} (b)) in the V-band AAVSO light curve using the Gaussian process regression method \citep[][]{celerite1, celerite2}\footnote{\url{https://celerite2.readthedocs.io/en/latest/}} to fit the main outburst. For the fitting prescription, we set the kernel function of the Gaussian process model to a mixture of {\tt SHOTerm} terms: one quasi-periodic component and one non-periodic component.

To analyse the time delay between the V-band and X-ray dips, we apply the Interpolated Cross-Correlation Function (ICCF) method, which has been used in time series analysis of BH-LMXBs and AGN reverberation mapping \citep[e.g.][]{bentz2009lick, 2023Sci...381..961Y}. Using the ICCF analysis code\footnote{\url{https://github.com/Bei-You-BH/MAD}}, we determine the the peak time delay $\tau_{p} = 11.00^{+1.18}_{-2.07}$ days and centroid delay $\tau_{c} = 10.40^{+1.18}_{-1.17}$ days. 
These delays are consistent within 1$\sigma$ uncertainties.

\begin{table}
\caption{\label{t1}Results of the power-law fitting ($F_\mathrm{\nu,opt} \propto {F_\mathrm{X}}^{\beta}$)  to the correlations between the optical ($g^{\prime}$ band) flux density and the 2-10 keV X-ray (\swift/XRT, \hxmt/LE) flux.}
\centering
\scalebox{0.9}{
\begin{tabular}{lllll}
\hline\hline\\
band & state &$\beta$ &$\rho$ &p-value\\
\hline\\
$g^{\prime}$--LE & HS1(decay) & $1.86\pm0.12$ & 0.85 & $5.12\mathrm{e}{-17}$\\
\hline\\
$g^{\prime}$--LE & SS& $0.58\pm0.03$& 0.88& $3.27\mathrm{e}{-12}$\\
\hline\\
$g^{\prime}$--LE & SS+IMS& $0.58\pm0.02$& 0.86& $1.42\mathrm{e}{-15}$\\
\hline\\
$g^{\prime}$--LE & HS2(optical rise)&  $-0.29\pm0.02$&  -0.93& $3.97\mathrm{e}{-05}$\\
\hline\\
\multirow{2}{*}{$g^{\prime}$--XRT} & HS1(rise)& \multirow{2}{*}{$0.45\pm0.01$} & \multirow{2}{*}{1.00~\tablefootmark{(a)}}& \multirow{2}{*}{$7.02\mathrm{e}{-25}$}\\
& +HS2(decay)&&&\\
\hline\\
$g^{\prime}$--XRT & SS& $0.50\pm0.05$& 0.71& $6.81\mathrm{e}{-04}$\\
\hline\\
$g^{\prime}$--XRT & SS+IMS& $0.54\pm0.03$& 0.62& $1.01\mathrm{e}{-04}$\\
\hline\\
$g^{\prime}$--XRT & mini-rises& $0.26\pm0.02$& 0.93& $3.00\mathrm{e}{-06}$\\
\hline\\
\multirow{3}{*}{$g^{\prime}$--XRT} & HS1(rise)& \multirow{3}{*}{$0.45\pm0.01$}& \multirow{3}{*}{0.98}& \multirow{3}{*}{$1.50\mathrm{e}{-42}$}\\
 & +HS2(decay)& & & \\
 & +mini-decays& & & \\
\hline
\end{tabular}}
\tablefoot{\tablefoottext{a}{The ranks of the X-ray and optical band data are identical (i.e., the data are sorted in the same order). In this case, the Spearman correlation formula necessarily gives a result of 1.}
}
\end{table}

\subsection{Optical / X-ray correlations}
\label{subsec: correlation}
X-ray monitoring provides better coverage than the optical band, therefore, we interpolate the observed $g^{\prime}$-band light curves (see Fig.~\ref{fig:interpolation_1}) 
using the interpolation method provided by python module \texttt{scipy.interpolate.CubicSpline}. This approach allows us to make full use of the X-ray data when analysing optical/X-ray correlations. We split the complete light curve into several segments in case the python routine is unable to obtain a smooth interpolation. 
After implementing the interpolation on the $g^{\prime}$-band light curve, we use the optical data derived from the resulting interpolation curve to plot correlations with \hxmt/LE and \swift/XRT data (see Fig.~\ref{fig:g-hxmt_correlation_interpo}, \ref{fig:g-xrt_correlation_interpo} and \ref{fig:g-xrt_correlation_fit}). In this study, we performed non-linear least squares fitting using the \texttt{curve\_fit} function from \texttt{SciPy}. We modelled potential correlations with a power-law function of the form, $F_{\nu, opt} \propto F_{X}^{\beta}$, following previous work \cite[e.g.][]{2006MNRAS.371.1334R, 2010MNRAS.402.2671R, 2013MNRAS.428.3083A, 2020MNRAS.493..940L, 2022ApJ...932...38S}, to derive slopes ($\beta$). For evaluating the significance of the correlation, we use the \texttt{spearmanr} function from \texttt{SciPy} to calculate the Spearman correlation coefficient, $\rho$, and corresponding p-value. The results are presented in Table \ref{t1}. All correlation diagrams show complex evolutionary tracks. 

Fig.~\ref{fig:g-hxmt_correlation_interpo} shows the optical/X-ray correlation of the main outburst of \SO, using the 2–10 keV X-ray fluxes derived from \hxmt\ spectral fitting and the $g^{\prime}$-band optical flux densities from LCO. The resulting slopes have distinct values in different stages.
During the first hard state, the source first evolves from the centre of the diagram towards the upper right corner, and then towards the lower left with a very steep slope of $\beta = 1.86\pm0.12$. We need to note that we only fit the points on the trajectory evolving towards the lower left (i.e. the decay phase of HS1) because of the complexity of the relation in the HS1. After that, there is a clockwise loop that corresponds to the period of dip event. Given the high-cadence observation in optical and X-ray observation during the period of dip event, the loop can be also seen in the non-interpolated optical/X-ray correlation diagrams (see Fig. \ref{fig:g-hxmt_nointerpo} and \ref{fig:hxmt_aavso_corr}). A similar loop is reported in the UV/X-ray flux correlation of Swift J1910.2--0546 \citep[see figure 5 in ][]{2014ApJ...784..122D}, although it is not clearly visible. We also fit the correlations in the SS ($\beta = 0.58\pm0.03$) and the optical rise phase ($\beta = -0.29\pm0.02$) of the HS2. 
We obtain a same slope ($\beta = 0.58\pm0.02$) as that in soft state when considering the soft state and intermediate state together.

Fig.~\ref{fig:g-xrt_correlation_interpo} displays the correlations between the 2--10 keV \swift/XRT X-ray fluxes and the $g^{\prime}$-band optical flux densities, revealing an extended correlation associated with the tail of the second hard state and three mini-outbursts. The upper right corner of Fig.~\ref{fig:g-xrt_correlation_interpo} (a) shows an evolution similar to that shown in Fig.~\ref{fig:g-hxmt_correlation_interpo}. We highlight three mini-outbursts in the lower-left corners of panels (b)--(d) in Fig.~\ref{fig:g-xrt_correlation_interpo}, where we find that the upper and lower tracks correspond to the rise and decay phases of the mini-outbursts, respectively. We fit the optical/X-ray correlation during the initial rise of the first hard state and decay of the second hard state, and obtain a strong correlation with a power-law index of $0.45\pm0.01$ (see Fig. \ref{fig:g-xrt_correlation_fit} (a)). During the soft and intermediate states, we obtain results consistent with those derived from the $g^{\prime}$/X-ray flux (\hxmt) correlation (see the fit in  Fig. \ref{fig:g-xrt_correlation_fit} (a)). In the evolution track of the first mini-outburst, it shows that optical brightening precedes the X-ray brightening during the the rising phase. 
For the optical/X-ray flux correlation in the three mini-outbursts (see the fit in Fig.~\ref{fig:g-xrt_correlation_fit} (b) and (c)), we fit the correlation during the decay phases of the three mini-outbursts and the rise of HS1 and the decay of HS2. A highly significant correlation is observed ($\rho$ = 0.98, p-value = $1.50\times10^{-42}$), characterised by a power-law index of 0.45$\pm$0.01. For the rising phases of the mini-outbursts, there is also a significant correlation ($\rho$ = 0.93, p-value = $3.00\times10^{-6}$) with a shallower power-law index of 0.26$\pm$0.02.

\section{Discussion}
\label{sec: Discussion}

We have investigated the evolution of long-term optical, UV and X-ray light curves of \SO\ throughout the main outburst and followed-by three mini-outbursts. 
We observed a broad dip event in both the X-ray and V bands before the transition from hard to soft state, with a delay of approximately 10 days between the two bands. 
We analysed the optical colour evolution in the colour-magnitude diagram. We found that, in general, the source becomes bluer as it brightens, and the colour in hard state is redder than in the soft state. The colour in the quasi-quiescence varies by $\sim$ 0.5 magnitudes.
We also found that, before the hard-to-soft state transition, the optical colour become stabilised earlier than the hardness ratio; and optical/UV ``bumps'' appears during and after the transition to hard state.
Finally, we analysed the correlations between the optical flux density and X-ray count rate during the main outburst and three mini-outbursts, which shows a complex correlations. In the correlation diagram (e.g., Fig. \ref{fig:g-hxmt_correlation_interpo}), we identified a distinct loop track before the transition to the soft state, which has appeared in a previous study \citep{2014ApJ...784..122D} but has never been observed with such clarity.

\subsection{Interpretation of the colour evolution}

From the global evolution of the $g^{\prime}-i^{\prime}$ colour in Sec. \ref{subsec: overall_colours}, it is apparent that in QQS the source is redder, and in outburst is bluer (see Fig. \ref{fig:cmd}). The general trend means that the disc in QQS is cooler if the majority of the optical emission originates from the accretion disc during the outburst and QQS \citep[][]{2008ApJ...688..537M, 2011MNRAS.416.2311R}. Additionally, we observe a wide range of optical colour during QQS, indicating that the source was not in a true quiescent state and other optical emission components were contributing in that stage. \citet{2023ATel16192....1B} already reported that quiescence was entered only in 2023, and its quiescent level is known to be slightly fainter than the QQS data in the CMD.

Comparing to the soft state, we find the $g^{\prime}-i^{\prime}$ colour in the hard state of the main outburst and in the mini-outburst is more variable. This variability may be attributed to the rapid jet flickering \citep[][]{2010MNRAS.404L..21C, 2017NatAs...1..859G, 2018ApJ...867..114B, 2019MNRAS.490L..62P, 2021MNRAS.505.3452P,2021MNRAS.504.3862T}.
On the one hand, the $g^{\prime}$ and $i^{\prime}$ data that we used to calculate the colour are within a bin size of 1 day, and rapid flickering usually occurs on timescales less than the exposure time (1 day) and filter switching, so the $g'$ and $i'$ data are not taken simultaneously. Therefore, the rapid jet flickering likely leads to the variable colour.
Baglio et al. (in preparation) also shows that optical variability in \SO\ in the hard state could be due to the jet. 
The hard state data are also redder than the soft state data, which supports the jet interpretation, since the jet component is redder than the disc component. This redder data in the CMD has also been reported in other sources \citep[][]{2011MNRAS.416.2311R, 2020ApJ...905...87B, 2023ApJ...949..104S} 

We see a drop and stabilisation of the $g^{\prime}-i^{\prime}$ colour preceding a dramatic change in spectral hardness, even before the transition to the intermediate state. The drop in the optical colour is probably due to the jet component starting to fade, which normally occurs around the same time as the start of the X-ray softening \citep[see e.g.][]{2010MNRAS.405.1759R, 2011A&A...534A.119C, 2012MNRAS.427L..11Y, 2013ApJ...768L..35R, 2014ApJ...784..122D,2018ApJ...867..114B}. However, in our case, the behaviour of optical colour indicates the jet component fades earlier than the X-ray softening, making this an unusual situation. A similar phenomenon is also observed in Swift J1910.2--0546 \citep[see the figure 3 in][]{2014ApJ...784..122D}.
In addition, the hot flow model predicts that the optical spectrum becomes harder when the hard-to-soft transition occurs, which is consistent with the bluer $g^{\prime}-i^{\prime}$ colour that we observed before the hard-to-soft state transition \citep[][]{2013MNRAS.430.3196V}.

There are ``colour bumps'' that occur during and after the transition to the soft state. The red ``colour bump'' that appears after the soft-to-hard transition is due to the recovery of the jet \citep[][]{2022MNRAS.514.3894O, 2024ApJ...962..116E} that makes energy spectrum redder. It is similar to other BH-LMXBs when they enter the hard state \citep[e.g.][]{2013ApJ...779...95K, 2013ApJ...768L..35R, 2014ApJ...795...74D, 2019MNRAS.482.2447P}. 
The smaller "bump" observed during IMS2 (MJD 58380.0–58393.0) is evident in the UV colours but not in the optical ($g^{\prime}-i^{\prime}$) colour, suggesting it is unlikely to be jet-related. 
Such a feature contradicts the expected jet behaviour, which would be more prominent in the optical band than in the UV. Moreover, jets typically do not appear in the soft state, though they can emerge during the soft-to-SIMS transition \citep[see][]{2020MNRAS.495..182R}. Further investigation through spectral energy distribution modelling is needed, but we here do not explore this in detail here.

\subsection{Optical and X-ray dips}
\label{subsec: discuss dip}

From the cross-correlation function analysis, we detect a centroid delay of $10.40^{+1.18}_{-1.17}$ days in the X-ray dip with respect to the optical dip (Fig. \ref{fig:HXMT-aavso}). A similar dip was observed in Swift J1910.2--0546 \citep[][]{2014ApJ...784..122D, 2023MNRAS.524.4543S}, with a delay of 6-7 days. The dip in Swift J1910.2--0546 also preceded the transition to the soft state, and the delay was interpreted as being due to the viscous time scale of the disc. In this scenario, the explanation for the observed delay is a mass-transfer instability that originates at the outer edge of the accretion disc and propagates inward. Based on this scenario, we propose that a decrease in the mass accretion rate in the disc leads to forming a ``gap'' in the matter density, followed by an increase. That gap then propagates on the viscous timescale, resulting in a UV dip followed by an X-ray dip, a hardness dip and a transition to the soft state. This process could be occurring in \SO, although the UV dip is not evident, probably due to the limited data coverage. We suggest that the enhanced mass transfer in the disc following the dip increases the matter density, causing the inner disc to move inward and initiating the state transition. We suggest that the propagating dip could be a potential tool for predicting impending state transitions, although it has been observed in only two sources. Future high-cadence optical-UV-X-ray monitoring of suitable sources is needed to help confirm the role of the propagating dip event in state transitions and determine whether it is a common phenomenon.

The collapse and recovery of the hot flow \citep[][]{2013MNRAS.430.3196V} could interpret the flux drop at longer wavelengths preceding the subsequent drop at shorter wavelengths, which seems consistent with the optical/X-ray dip observed in Swift J1910.2--0546. However, this model requires that the UV emission rises before near-infrared emission when the source transitions back to soft state, which is the opposite of the case observed in Swift J1910.2--0546 \citep[][]{2014ApJ...784..122D, 2023MNRAS.524.4543S}. In \SO, the optical rise part of the dip occurs $\sim$ 10 days before the X-ray rise. The delay between UV and optical dips is less well constrained, but optical rise also appears to occur before UV rise  (see Fig. \ref{fig:time_evol_2}).
Moreover, the optical colour is bluer (i.e. the drop of $g^{\prime}-i^{\prime}$ colour) in the rise of the dip, indicating that it is the disc that is brightening, not the redder component (i.e. jet or hot flow). Therefore, collapse and recovery of the hot flow is not the cause of the optical/X-ray dip. 

Interestingly, we noticed that during the optical dip event, the superhump modulation reported by \citet{2022MNRAS.509.1062T}. were detected, beginning near the optical decline reversal. This modulation exhibits the largest amplitude observed in black hole X-ray binaries. They suggest that such large-amplitude modulation is caused by the disc warping, driven by irradiation. As we proposed above, the dip is likely due to the ``gap'' in the density in the disc. Therefore, we speculate that this warp may be related to the presence of the gap. Assuming a gap would form in the disc, causing a optical fading. Meanwhile, if the matter density in the gap is uneven and it might induce a warp, which is formed near the minimum of the optical dip, then as a result, the large-amplitude optical modulation emerges. Subsequently, due to the enhanced accretion rate, more accreting material from the outer disc flows in, filling the gap. This process could also be uneven, sustaining the warped disc. Consequently, the large-amplitude optical modulation persists while the optical flux is increasing. 
Eventually, the gap vanishes, and the source transitions to the soft state. The disc then reaches a more stable, steady state. During this phase, the disc warping diminishes but remains present, as indicated by the comparable large modulation measured by \citet{2022MNRAS.509.1062T}.
Overall, the emergence of disc warping in \SO\ could be related to the formation of the dip event, and lasts longer than the duration of the dip events.

\subsection{Implication of optical / X-ray correlations}
\label{subsec: discuss correlation}
Investigating the optical/X-ray correlation (See Fig. \ref{fig:g-hxmt_correlation_interpo} and \ref{fig:g-xrt_correlation_interpo}), there are two main correlation tracks observed. 
One corresponds to the initial rise and final decay phases of the hard state ($\beta = 0.45\pm0.01$). The other track is in the soft state and intermediate state ($\beta = 0.58\pm0.02$), as shown in the fit in Fig. \ref{fig:g-xrt_correlation_fit} (a). We found that, for a given X-ray flux in \SO, the optical flux is stronger in the HS than in the SS, which may indicate a contribution from a jet to the optical emission during the HS. 
The double track was also observed in other source, such as XTE J1550--564, 4U 1543--47 \citep[][]{2007MNRAS.379.1401R} and GX 339--4 \citep[][]{2009MNRAS.400..123C}.
Below we will discuss the optical/X-ray correlation of \SO\ in different states.

During the initial rise HS1 and final decay of the HS2, both period follows same correlation (see Fig. \ref{fig:g-xrt_correlation_interpo} (a)), with a strong correlation characterised by a power-law index of $0.45\pm0.01$. In both period, the optical emission could be dominated by the X-ray reprocessing \citep[][$\beta\approx0.5$]{1994A&A...290..133V}, though the viscous disc contributions \citep[][$\beta\approx0.3$]{2006MNRAS.371.1334R} may slightly reduces the slope below pure X-ray reprocessing predictions. 
Broadband spectral modelling \citep{{2024ApJ...962..116E}}confirms significant jet contributions to optical emission. 
However, our the optical/X-ray correlation index is not consistent with the theoretical value of jet model \citep[][$\beta\approx0.7$]{2006MNRAS.371.1334R}, which suggests that the jet emission is not the dominant component in the optical band.

During the decay of HS1, as shown in Fig. \ref{fig:g-hxmt_correlation_interpo}, it follows a very steep slope of the optical/X-ray correlation, $\beta = 1.86\pm0.12$, which is even higher than the prediction of the jet.
The study of broad spectral energy distributions in previous work shows that the jet spectral break decreases in frequency by three orders of magnitude during the decay phase of HS1 \citep[][see figure 2 (a) and figure 3 (a)]{2024ApJ...962..116E}. Therefore, one possible interpretation for such a steep slope is that the jet spectral break is shifting down in frequency. This would cause an optical drop as the higher frequency emission from the jet fades, without fading in the X-ray or radio. 

Before the transition to the soft state,
there is a loop-like track in the plot of Fig. \ref{fig:g-hxmt_correlation_interpo} (corresponding to the dip period), which has never been reported in the optical/X-ray correlation analysis of any other source. The detection of this loop in the optical/X-ray correlations is seen from the optical (LCO and AAVSO) and X-ray (\hxmt) observations before the transition to the soft state. The loop track appears to emerge in UVOT/XRT correlation \citep[see figure 5 in ][]{2020ApJ...889..142S} but not clear due to the lack of sufficient observation from UVOT during that period. 
This loop was also observed in Swift J1910.2--0546 \citep[see figure 5 in][]{2014ApJ...784..122D}, where a delayed X-ray dip relative to the optical/UV dip is detected. However, it is not as clear as in our case due to the lower cadence of observation in their study.
The loop indicates that the optical brightening precedes the X-ray brightening. We suggest that it is related to an increase in mass accretion rate in the outer disc. The material then propagates inward through the disc to the X-ray emitting region, causing the X-ray brightening and triggering the transition to the soft state, as our discussion of the rise phase of the dip in Sec. \ref{subsec: discuss dip}. It is worth mentioning that the discovery of this loop structure is due to the high-cadence observations in the optical (LCO and AAVSO) and X-ray bands (\hxmt). Conducting high-cadence, multi-wavelength observations during the outburst in the future may help us identify similar loop-like structures in other sources.

During the soft state, the optical/X-ray correlation shows a power-law index of $0.58\pm0.03$ as the source fades (see Fig. \ref{fig:g-hxmt_correlation_interpo}).
A combined fit of the soft state and intermediate state yields an identical slope ($\beta = 0.58\pm0.02$; Fig. \ref{fig:g-hxmt_correlation_interpo}). Both values are marginally higher than standard X-ray reprocessing predictions. Recent studies demonstrate that the expected slope ($\beta$) of the optical/X-ray correlation depends on the heating mechanisms (viscous vs. irradiated disc) and the spectral region of the disc emission, spanning from the Rayleigh-Jeans (RJ) tail to the multi-colour blackbody plateau
 \citep{2009MNRAS.400..123C, 2020MNRAS.495.3666T,2015MNRAS.453.3461S}. For a viscously heated disc in the soft state, $\beta$ ranges from 0.26 (RJ-dominated) to 0.67 (disc-dominated). In contrast, X-ray reprocessing produces a broader range of 0.28 $\leq \beta \leq$ 1.33. Besides, the colour-magnitude diagram (see Fig. \ref{fig:cmd}) shows the soft-state data following what is expected for the optical emission from disc. However, the measured $\beta \approx 0.58$ overlaps with predictions from both mechanisms, making it ambiguous to distinguish their contributions. Broadband spectral modelling resolves this ambiguity: the optical data in the soft state are best described by an irradiated disc \citep[][]{2024ApJ...962..116E}, strongly suggesting X-ray reprocessing dominates the optical emission.

During the three mini-outbursts, their rise phases follow a distinct track separate from the decay phases, with the former on the upper track and the latter on the lower track (see Fig. \ref{fig:g-xrt_correlation_interpo}(c)--(d)). On the one hand, this means that different dominant mechanisms contribute to the optical emission of the rise and decay phases, respectively.
\citet{2020ApJ...889..142S} also displayed a similar correlation using the UVOT/XRT data, but only has a single track for mini-outbursts, likely due to the lack of data during the rise phases.
The $\beta$ in the rises of mini-outbursts is $0.26\pm 0.02$ (see Fig. \ref{fig:g-xrt_correlation_fit}(b)), which is consistent with the viscously heated accretion disc \citep[][]{2006MNRAS.371.1334R}.
We also notice that the rise of HS1 and the decay of HS2 align with the correlation in the decays of mini-outbursts, suggesting that the same process is responsible for producing the optical emission. We fit them together and obtain a nice optical/X-ray correlation in the decay phases with $\beta = 0.45 \pm 0.01$ (see Fig. \ref{fig:g-xrt_correlation_fit}(c)), which suggests that the optical emission is dominated by the X-ray reprocessing.
On the other hand, these two separated tracks in the first mini-outburst also indicate that the optical brightening might precede the X-ray brightening. \citet{2021MNRAS.504.4226S} also estimated a time delay of 20 days between the B band and XRT ($2-10$ keV) flux in the first mini-outburst. This is expected as the heating front travels through the disc during the initial rise \citep[e.g.][]{2001A&A...373..251D}. 

\section{Conclusions}

In this work, we present the analysis of optical, UV and X-ray observations of the main outburst and the subsequent mini-outbursts of the black hole X-ray binary \SO. Our main results and conclusions are:

\begin{enumerate}

\item We detect optical and X-ray flux dips before the hard-to-soft transition in the main outburst. There is a $\sim$ 10-day delay between the X-ray and the optical dips. We suggest that it reflects the time it takes, on the viscous timescale, for a decrease then increase of mass density to propagate from the outer disc to the inner disc. The transition to soft state is likely due to the enhanced mass transfer in the disc following the dip. The propagating dip might be a possible tool for predicting the impending state transition.

\item We observe that the optical colour becomes bluer and less variable before transition to intermediate state, likely due to quenching of the variable jet. Typically, the optical colour is expected to be bluer at the same time as the X-ray spectral softening during the transition to the soft state. However, this prominent change in optical colour even occurs during the hard state, indicating that our finding is an unusual case.

\item We measure the steepest optical/X-ray correlation slope in a BH-LMXBs to date during the decay phase of the first hard state. This might be related to the jet spectral break shifting to a lower frequency.

\item We find a loop track in the optical/X-ray flux correlation diagram before the transition to the soft state. This is caused by the time delay between the X-ray and optical dips. The detection of the optical/X-ray dip events and the loop track is revealed through the high-cadence monitoring in the optical and X-ray band. Conducting the high-cadence, multi-wavelength monitoring could help us further understand the role of dip event in the state transition.

\item We find that the rise and decay phases of mini-outbursts follow two distinct tracks, indicating that different mechanisms dominate the optical emission in each phase. The optical rising phase appears governed by viscous heating in the accretion disc, while the fading phase is likely dominated by X-ray reprocessing in the disc.

\end{enumerate}

\begin{acknowledgements}
We thank an anonymous referee for useful comments that helped
us improve the paper. We thank Thabet Al Qaissieh and Aldrin Gabuya for providing the optical data from the Al Sadeem Observatory, Abu Dhabi, UAE. 
DMR and PS are supported by Tamkeen under the NYU Abu Dhabi Research Institute grant CASS. 
GB acknowledges support from the China Manned Space Program with grant No. CMS-CSST-2025-A13. 
PY acknowledges support from the China Scholarship Council (CSC), No. 202304910059.
\end{acknowledgements}

%\bibliographystyle{aa} % style aa.bst
\bibliographystyle{aa}
\bibliography{references.bib} % your references Yourfile.bib

\begin{appendix}
\section{EXTRA MATERIAL}

\begin{figure*}
\centering
    \includegraphics[width=1.75\columnwidth]{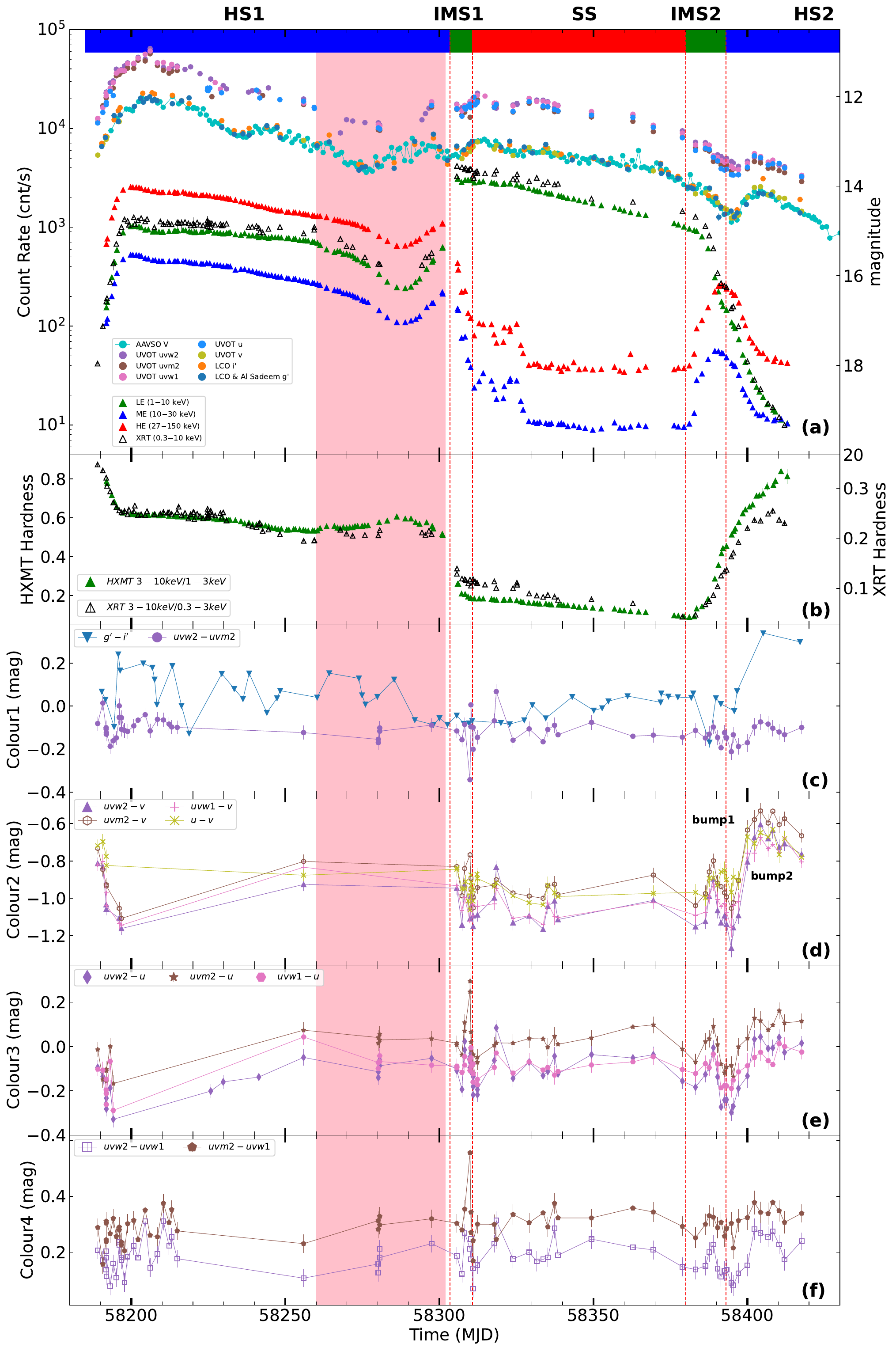}
    \caption{Multi-wavelength light curves, X-ray hardness and optical/UV colours during the main outburst of \SO.}
    \label{fig:time_evol_2}
\end{figure*}

\begin{figure*}
\centering
	\includegraphics[angle=90, width=1.3\columnwidth]{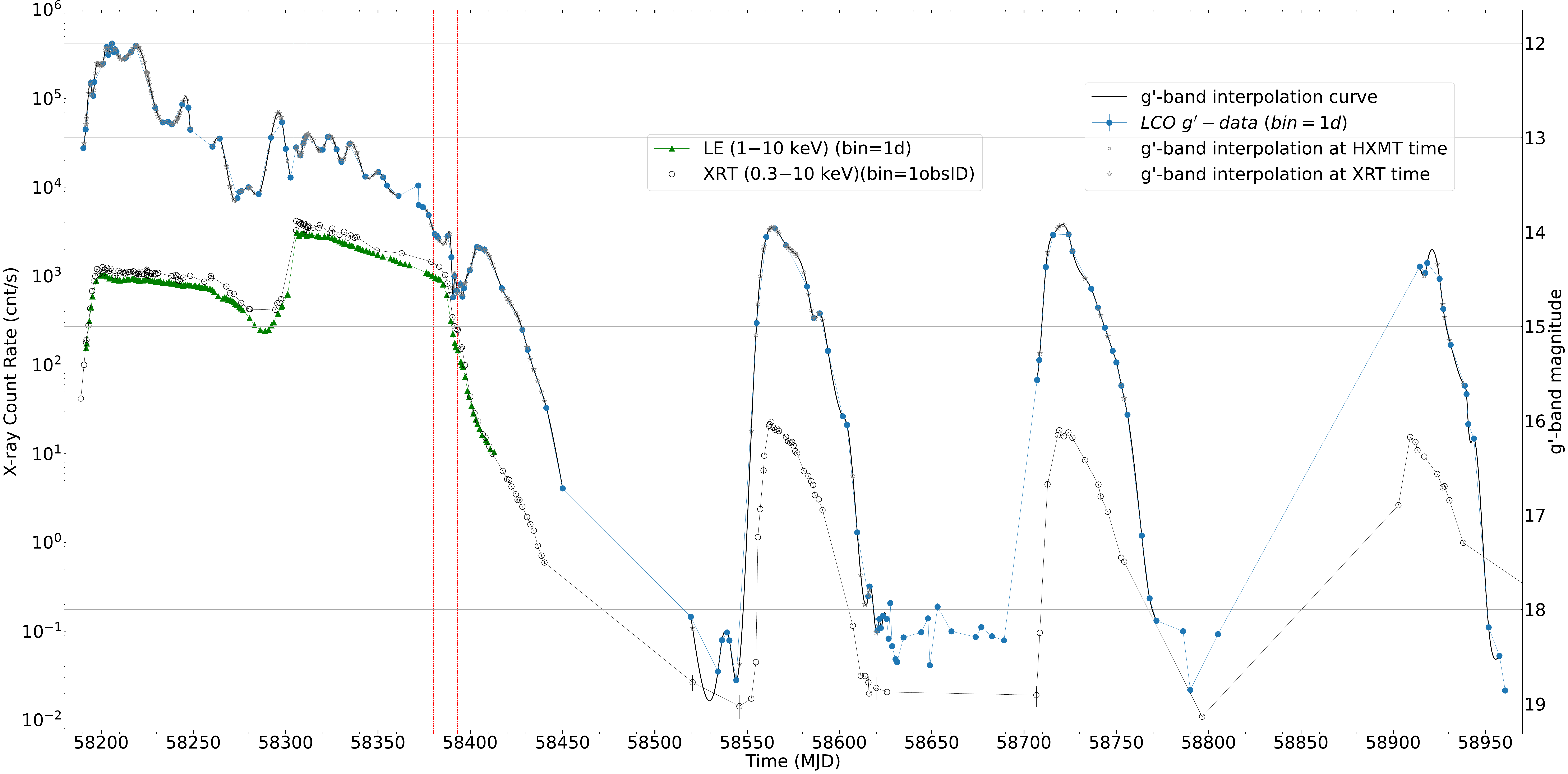}
    \caption{Interpolation on $g^{\prime}$ band light curves of \SO. Gray circles and gray stars are the interpolation at the \hxmt\ and XRT observation time, respectively.}
    \label{fig:interpolation_1}
\centering
\end{figure*}

\begin{figure}
\centering
    \mbox{\includegraphics[width=\columnwidth]{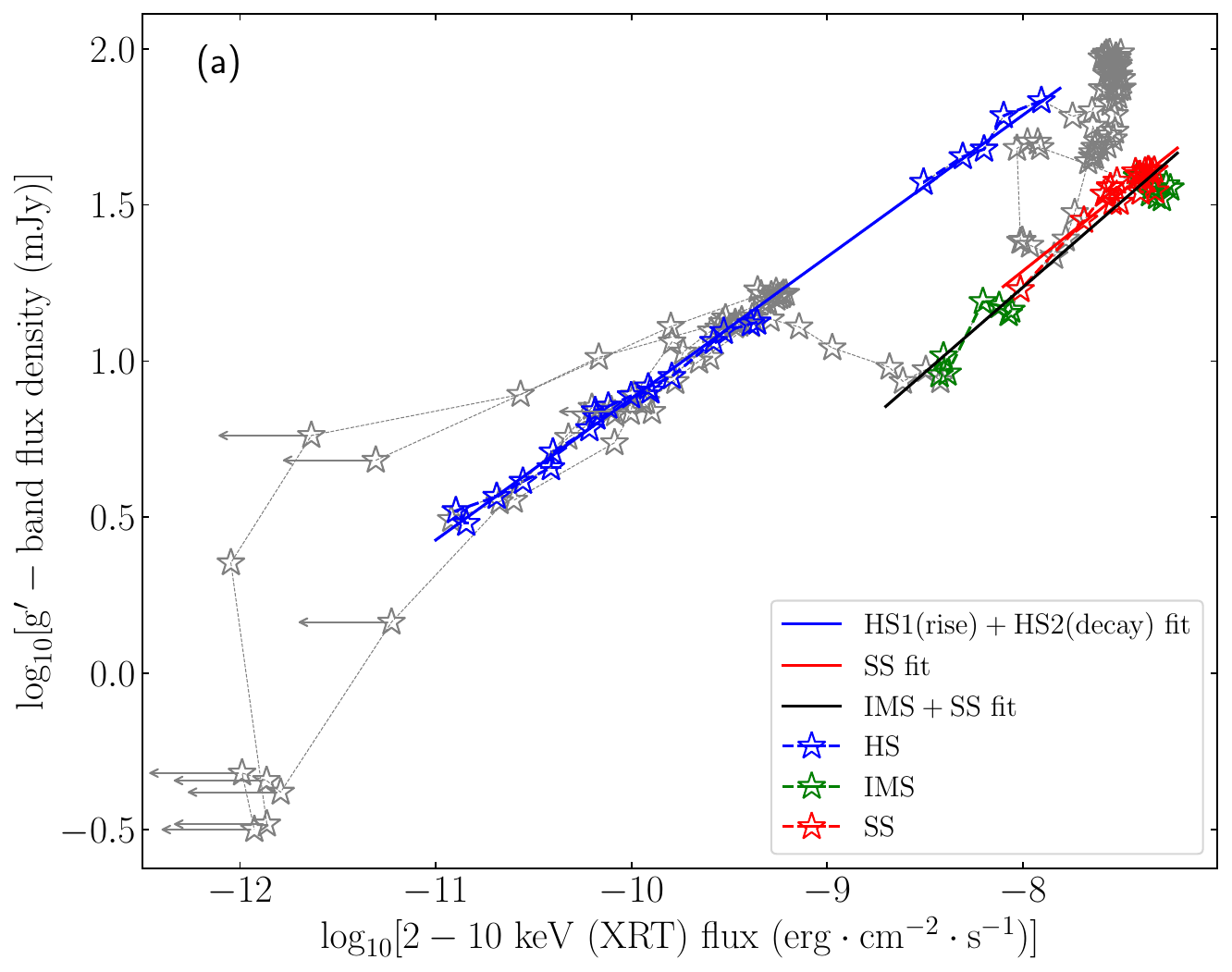}}
    \mbox{\includegraphics[width=\columnwidth]{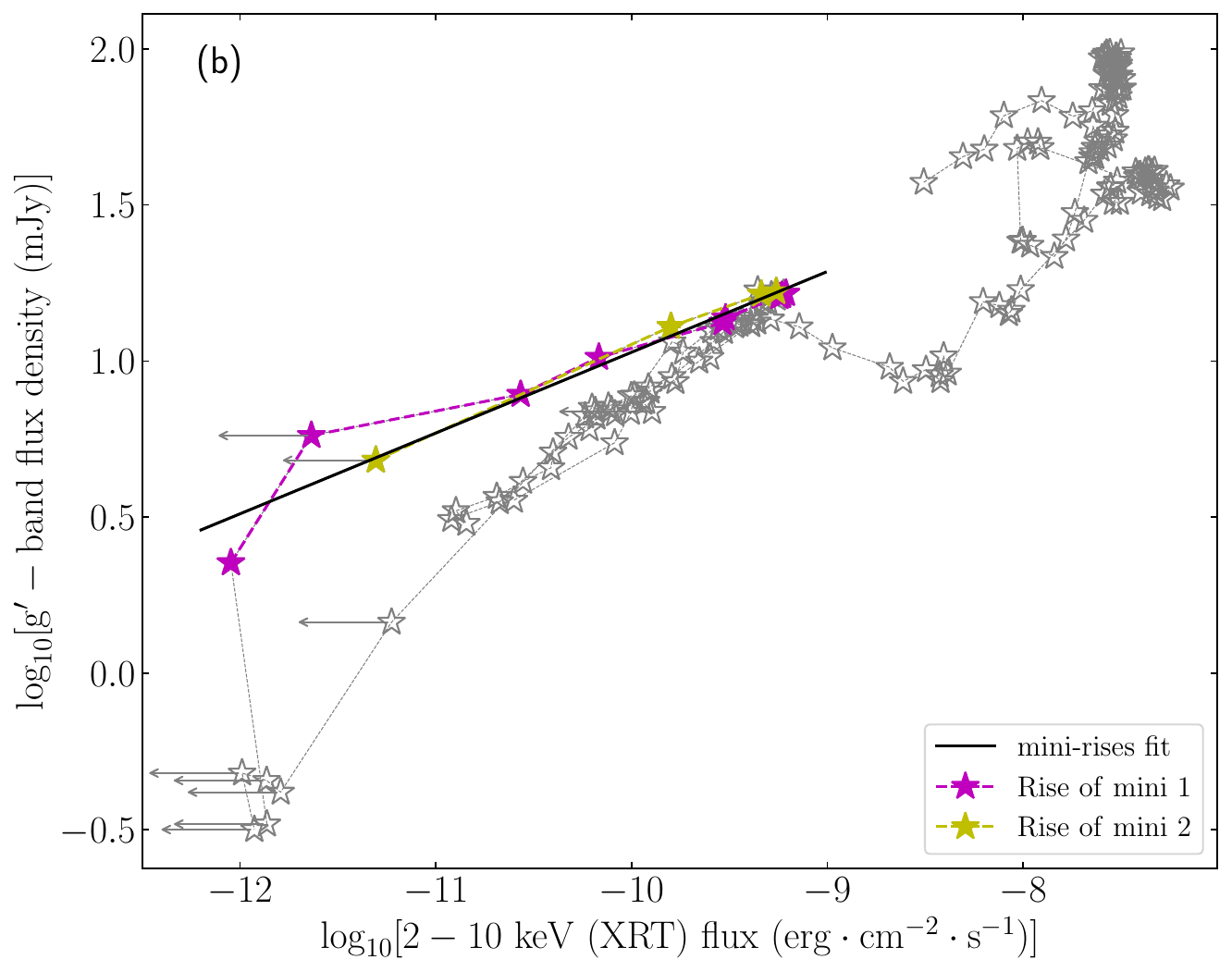}}
        \mbox{\includegraphics[width=\columnwidth]{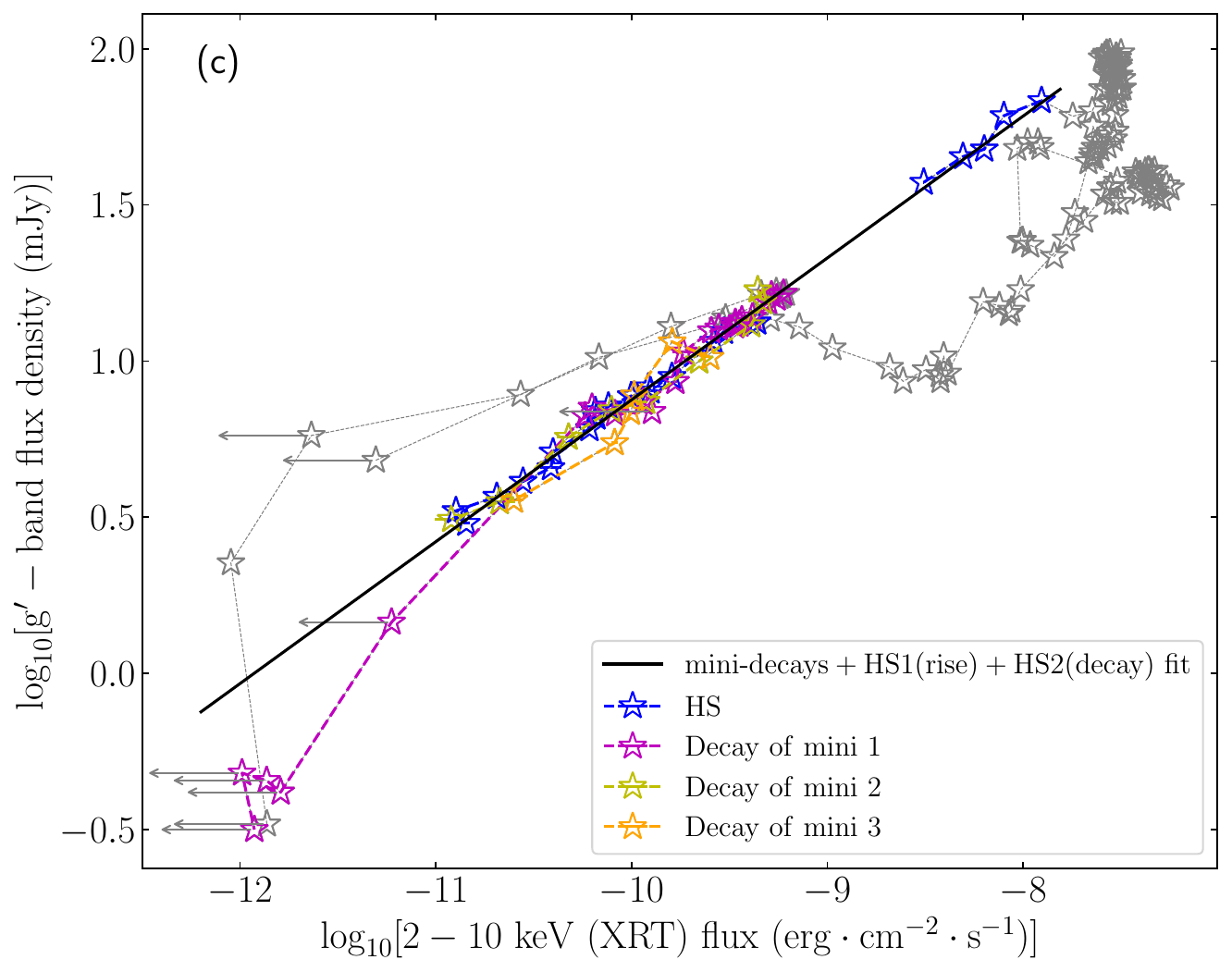}}
    \caption{Power-law fits to the correlations between the 2--10 keV \swift/XRT X-ray flux and the interpolated $g^{\prime}$-band flux density of \SO. 
(a): Joint fit to the initial rise of HS1 and the final decay of HS2, and a separate fit to the IMS and SS phases.
(b): Joint fit to the rise phases of mini-outbursts 1 and 2. 
(c): Joint fit to the initial rise of HS1, the final decay of HS2, and the decay phases of mini-outbursts 1, 2, and 3. \textbf{Note}: The upper-limit points are not included in the fitting of (c).}
    \label{fig:g-xrt_correlation_fit}
\end{figure}

\begin{figure}
\center
    \mbox{\includegraphics[width=\columnwidth]{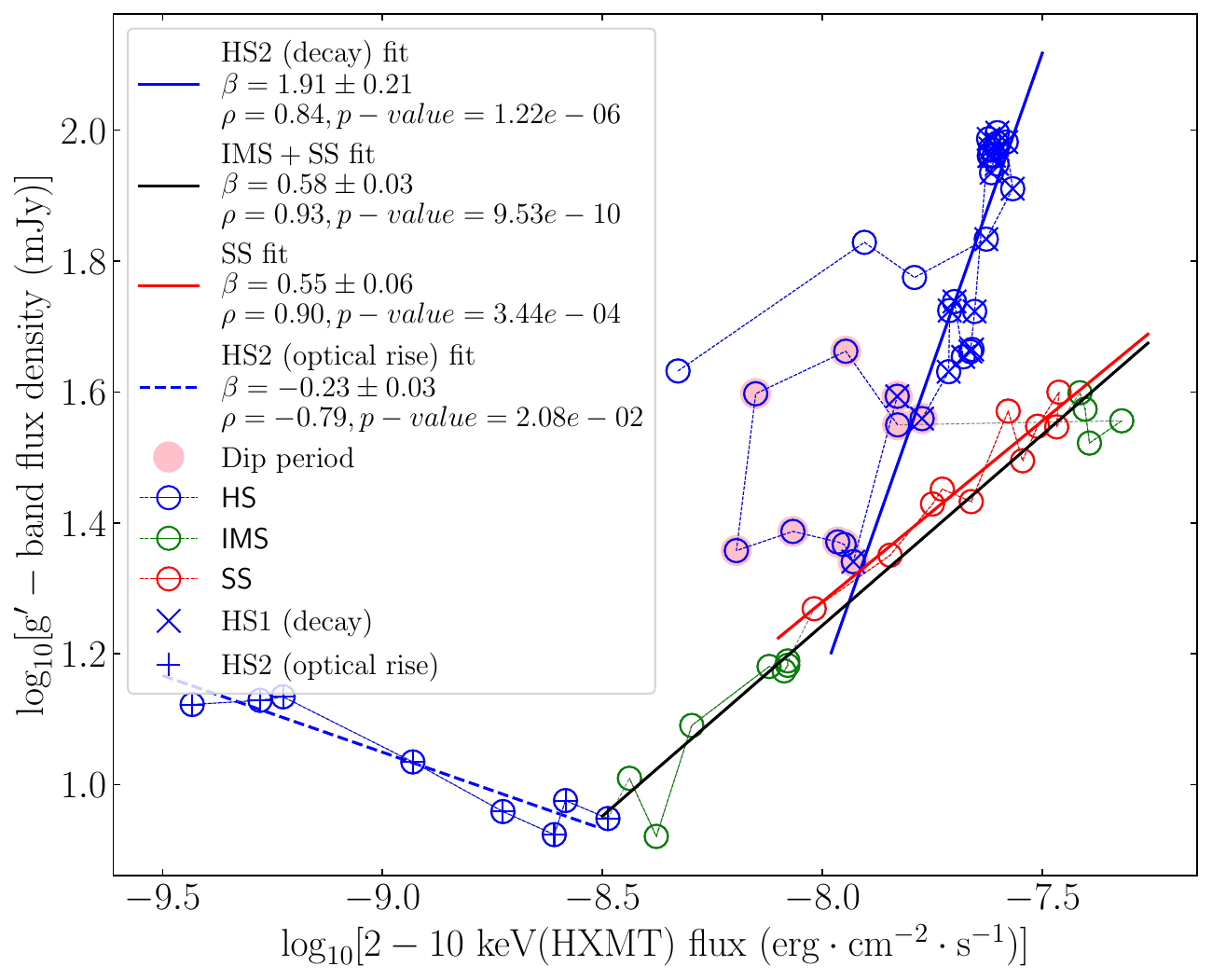}}
    \caption{Power-law fits to the correlations between the  2-10 keV X-ray (\hxmt) flux and the non-interpolated optical ($g^{\prime}$ band) flux density during the main outburst in \SO. }
    \label{fig:g-hxmt_nointerpo}
\end{figure}

\begin{figure}
\center
    \mbox{\includegraphics[width=\columnwidth]{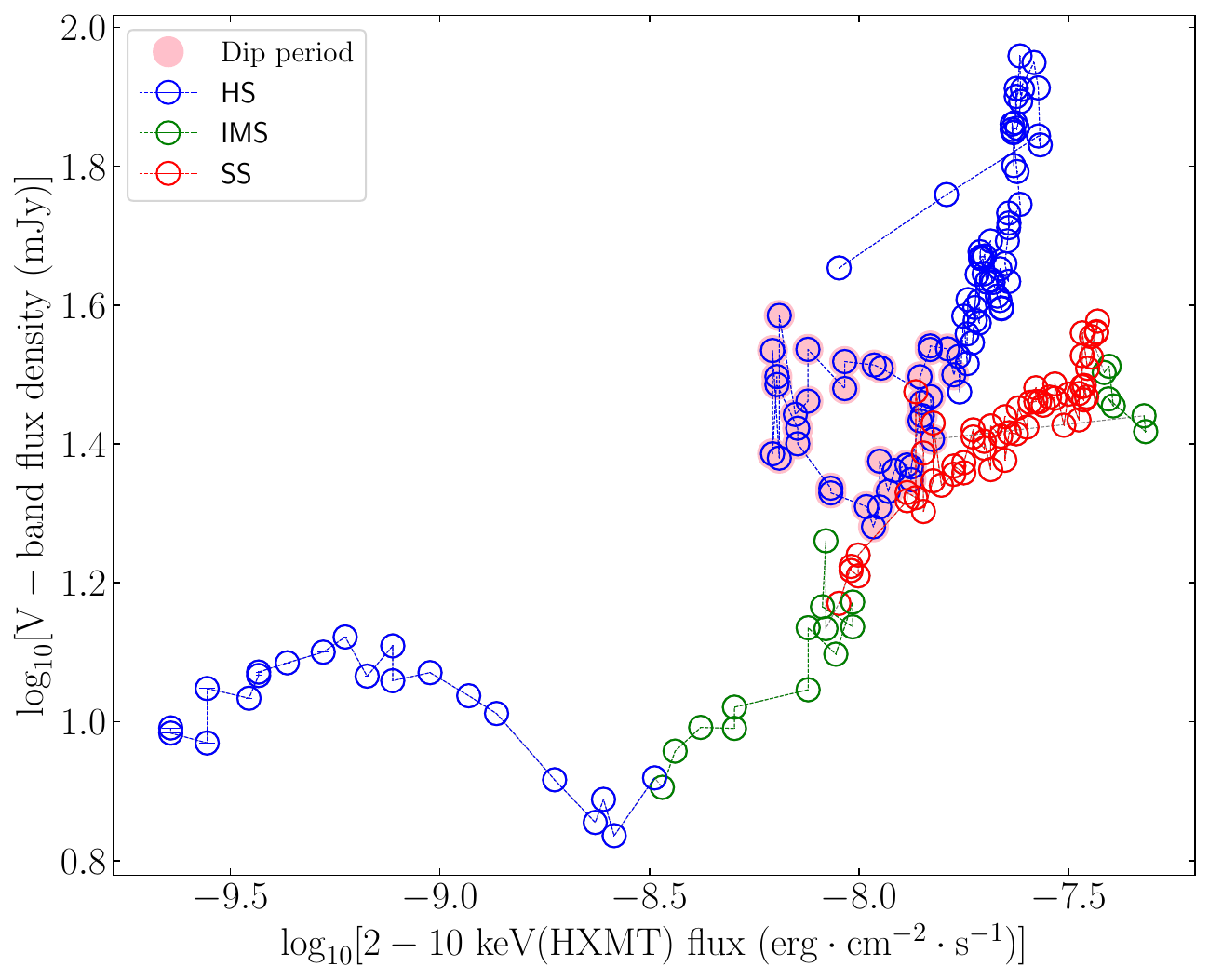}}
    \caption{Correlations between the 2-10 keV X-ray (\hxmt) flux and the optical (V band) flux density during the main outburst in \SO.}
    \label{fig:hxmt_aavso_corr}
\end{figure}

\end{appendix}

\end{document}